\documentclass[12pt]{article}

\usepackage[T1]{fontenc}
\usepackage{mathptmx}
\usepackage{amsmath,amssymb,amsfonts}
\usepackage[numbers,sort&compress]{natbib}
\usepackage{dsfont}
\usepackage{xcolor}
\usepackage{gensymb}
\usepackage{rotating}
\usepackage{subcaption}
\usepackage{caption}
\usepackage{booktabs}
\usepackage{tabularx}
\usepackage{array}
\usepackage{placeins}
\usepackage[hidelinks]{hyperref}
\usepackage[shortlabels]{enumitem}
\usepackage{graphicx}
\usepackage{tikz}
\usetikzlibrary{arrows.meta,positioning}
\usepackage[margin=1in]{geometry}
\usepackage{amsthm}

\usepackage{libertine}
\usepackage{cleveref}
\usepackage[all,defaultlines=3]{nowidow}

\crefname{lemma}{lemma}{lemmas}
\Crefname{lemma}{Lemma}{Lemmas}
\crefname{theorem}{theorem}{theorems}
\Crefname{theorem}{Theorem}{Theorems}
\crefname{proposition}{proposition}{propositions}
\Crefname{proposition}{Proposition}{Propositions}
\crefname{corollary}{corollary}{corollaries}
\Crefname{corollary}{Corollary}{Corollaries}
\crefname{remark}{remark}{remarks}
\Crefname{remark}{Remark}{Remarks}
\crefname{assumption}{assumption}{assumptions}
\Crefname{assumption}{Assumption}{Assumptions}
\crefname{section}{Section}{Sections}
\Crefname{section}{Section}{Sections}
\crefname{subsection}{Section}{Sections}
\Crefname{subsection}{Section}{Sections}
\crefname{subsubsection}{Section}{Sections}
\Crefname{subsubsection}{Section}{Sections}
\crefname{equation}{Equation}{Equations}
\Crefname{equation}{Equation}{Equations}
\crefname{algorithm}{Algorithm}{Algorithms}
\Crefname{algorithm}{Algorithm}{Algorithms}
\crefname{figure}{Figure}{Figures}
\Crefname{figure}{Figure}{Figures}
\crefname{table}{Table}{Tables}
\Crefname{table}{Table}{Tables}
\crefname{appendix}{Appendix}{Appendices}
\Crefname{appendix}{Appendix}{Appendices}

\usepackage{threeparttable}
\usepackage{longtable}
\usepackage{makecell}
\usepackage{multicol}
\usepackage{multirow}
\usepackage{siunitx}
\usepackage{bm}
\usepackage{float}
\usepackage{wrapfig}
\usepackage{algorithm}
\usepackage{algorithmic}
\usepackage{etoolbox}
\usepackage[normalem]{ulem}

\BeforeBeginEnvironment{wrapfigure}{\FloatBarrier\newpage}
\AtBeginEnvironment{algorithmic}{\scriptsize}

\newcolumntype{C}[1]{>{\centering\arraybackslash}m{#1}}
\newcolumntype{L}[1]{>{\raggedright\arraybackslash}m{#1}}

\DeclareMathAlphabet{\pazocal}{OMS}{zplm}{m}{n}

\title{Fixed-Dimensional Latent Flow for Generating Variable-Size 3D Molecules}

\author{
\small
\begin{tabular}{@{}cccc@{}}
Weichi Yao$^{1}$ & Cameron Gruich$^{2}$ & Bryan R. Goldsmith$^{2}$ & Yixin Wang$^{3}$
\end{tabular}\\[0.75em]
\small $^1$Michigan Institute for Data \& AI in Society\\
\small $^2$Department of Chemical Engineering\qquad
$^3$Department of Statistics\\
\small University of Michigan, Ann Arbor, USA\\
\footnotesize\texttt{\{weichiy,cgruich,bgoldsmith,yixinw\}@umich.edu}
}

\date{}

\begin{document}

\maketitle

\begin{abstract}
Molecular size is intrinsically linked to composition, structure, and properties, but most 3D molecular generators treat size as a predefined input rather than a quantity to be generated. We introduce Equivariant-Free Transformer-Autoencoded Latent Flow Matching, a two-stage framework that samples a fixed-dimensional latent vector using flow matching and employs an autoregressive Transformer to determine molecular size, atom types, coordinates, and chemical attributes. Canonical atom ordering and rigid-pose alignment enable Transformers without equivariant layers, while decoded attributes guide bond reconstruction. On PCQM4Mv2, unconditional generation yields 87.9\% unique and novel molecules passing sanitization and PoseBusters checks, exceeding baselines with lower end-to-end training and sampling time and higher end-to-end throughput. Across ten target HOMO-LUMO gaps, internal ranking retains 30\% of the screened candidates and increases the density functional theory-verified hit rate within 0.1 eV of the target from 25.0\% to 52.4\%, while largely preserving novelty and diversity. These results demonstrate fixed-dimensional latent generation with autoregressive decoding as a practical approach to molecular design without prespecifying size.
\end{abstract}

\section{Introduction}
Generative models could accelerate molecular discovery by proposing new molecules with three-dimensional (3D) structures. 
Molecule size, composition and geometry are interdependent. 
The number and types of atoms constrain the structures a molecule can adopt and the properties it can exhibit. 
Learning their joint distribution could help capture these dependencies across molecular sizes and thereby improve both exploration of chemical space and generation towards desired properties.
This motivates a generator in which molecular size emerges alongside composition and geometry, without a prespecified atom count.

Many 3D molecular generators, however, require the molecule size $N$ to be determined before the atom-wise generative process begins.
Molecule-space diffusion and flow models initialize $N$ atom states before beginning the generative process, and bond-explicit variants may additionally initialize $O(N^2)$ pairwise states~\citep{hoogeboom2022equivariant,vignac2023midi,hua2024mudiff,le2024eqgatdiff,dunn2024flowmol,irwin2025semlaflow}.
Some latent approaches also require $N$ in advance, as they generate one latent vector per atom for subsequent decoding~\citep{xu2023geoldm,joshi2025adit,luo2025uae3d}.
Other approaches use fixed-dimensional molecular representations, with decoders that either require $N$ specified in advance~\citep{li2026moldiffdae,chen2025molflae} or infer it from a generated molecular field~\citep{kirchmeyer2024funcmol}.
These considerations motivate our central architectural design, learning a single fixed-dimensional molecule-level latent vector $\mathbf z\in\mathbb R^d$ that can be decoded into a molecule without requiring $N$ to be specified in advance.

Turning a fixed-dimensional latent vector into a molecule of unknown size requires a decoder that can determine when construction is complete.
One latent approach handles variable size by growing an atom-wise latent sequence, using conditional diffusion for each new latent and a separate classifier for termination~\citep{ottomano2026kronos}.
Direct autoregressive 3D generators address variable size by growing molecules atom by atom~\citep{cheng2025quetzal,li2025inertialar,rose2026neat}.
For our latent generation design, we adopt this atom-by-atom strategy by using an autoregressive decoder that stops when it emits an end-of-molecule (EOM) token.

Autoregressive decoding faces ambiguity in its sequence and coordinate targets because the same molecule can be represented with different atom orders, positions, and orientations.
Existing approaches handle geometric symmetry through equivariant layers or $\mathrm{SE}(3)$ augmentation~\citep{hoogeboom2022equivariant,xu2023geoldm,joshi2025adit,cheng2025quetzal}.
Other approaches use learned coordinate frames~\citep{guo2025framedi} or learned rotational alignment~\citep{ding2025radm}, or incorporate symmetry into the generative process or training objective~\citep{ko2026permsymdiff,xu2026quotientdiffusion}.
Canonical approaches instead select a consistent representation before learning, for example, through inertial-frame tokenization or geometric spectral canonicalization~\citep{li2025inertialar,zhou2026canonicalization}.
We adopt a lightweight version of this representation-level strategy. Canonical graph ordering provides a reproducible atom sequence, while atom position alignment removes translational and proper-rotational symmetries to provide a reproducible coordinate frame.
For nondegenerate geometries, alignment preserves geometry and keeps enantiomers distinct by using translations and proper rotations, avoiding the ambiguity of reflection-invariant representations~\citep{dumitrescu2025chirality}.
Applied once before encoding, these deterministic preprocessing steps enable standard Transformer and feed-forward backbones.

A second challenge is to generate a chemically complete 3D molecular graph.
Some methods first generate a molecular graph and then predict its 3D coordinates~\citep{liu2025nextmol}.
Other 3D generators explicitly predict a dense $O(N^2)$ array of bond types, requiring consistency between the predicted bonds and atomic valences~\citep{huang2023jodo,dunn2025flowmol3,reidenbach2026megalodon,luo2025uae3d,li2026moldiffdae}.
Bond-free generative models generate atom types and Cartesian coordinates, leaving connectivity to postprocessing~\citep{hoogeboom2022equivariant,xu2023geoldm,joshi2025adit,cheng2025quetzal}.
Atom types and coordinates alone may be insufficient to reconstruct a fully specified molecular graph, particularly for heavy-atom scaffolds with unspecified hydrogen counts.
We take an intermediate route, where the generative model remains bond-free but augments atom types and Cartesian coordinates with additional atom- and molecule-level attributes to guide deterministic bond reconstruction without requiring a learned dense bond-prediction head.

A further challenge is to generate molecules with desired properties.
Existing diffusion and flow models incorporate target molecular properties as conditioning inputs~\citep{hoogeboom2022equivariant,xu2023geoldm,huang2023jodo,zeng2026propmolflow}.
Latent approaches either condition only the latent generator~\citep{kim2025g2ddiff,qiu2026driftingmol,zhang2025gcldm} or also condition the molecular encoder and decoder~\citep{lim2018molecular,kang2019conditional,xu2023geoldm,you2024latent}.
Property-supervised representations have also been used for latent optimization, predictor guidance, or targeted embedding edits~\citep{gomez2018automatic,lobo2026moltenflow,li2026moldiffdae}.
In our design, only the latent generator receives target properties as conditioning inputs, allowing unconditional and property-conditioned generation to use the same autoencoding architecture.
We optionally train the autoencoder to predict molecular properties and use these predictions to rank and filter candidates by how closely their predicted properties match the target. 

Together, these design choices define \emph{Equivariant-Free Transformer-Autoencoded Latent Flow Matching} (EF-TALFM), a practical framework for generating 3D molecular graphs without prespecifying the atom count. 
On PCQM4Mv2~\citep{hu2021ogblsc}, EF-TALFM achieves a higher yield of unique and novel molecules passing sanitization and PoseBusters~\citep{Buttenschoen2024posebuster} checks than the evaluated baselines, with lower end-to-end training and sampling times under the reported budgets.
For property-conditioned generation, internal-readout selection improves property-targeting accuracy while largely preserving structural novelty and diversity. 

\paragraph{Contributions.} Our contributions follow the architectural hierarchy above:
\begin{enumerate}[leftmargin=*]
    \item \textbf{Fixed-dimensional molecule-level generation.} We introduce a two-stage 3D molecular generator whose second-stage latent flow samples a single molecule-level vector $\mathbf z\in\mathbb R^d$, independent of molecular size.
    \item \textbf{Variable-size autoregressive decoding.} We decode the sampled latent vector with a causal Transformer that autoregressively generates atom types, 3D coordinates, and additional chemical attributes until an EOM token is generated.
    The decoder thus determines molecular size during generation, without a prespecified molecule size.
    \item \textbf{Representation-level symmetry handling.} We use canonical atom ordering and align each molecule to a standardized coordinate frame; for nondegenerate geometries, alignment preserves geometry and chirality. This deterministic preprocessing enables standard Transformer and feed-forward backbones without equivariant layers or rotational augmentation.
    \item \textbf{Bond-free chemical realization.} We generate enriched atom- and molecule-level chemical attributes that guide deterministic graph reconstruction without learning a dense bond-prediction head.
    Lightweight restrained geometry refinement further improves local geometric validity.
    \item \textbf{Property-conditioned generation and internal-readout selection.}
    We add a property-prediction objective during autoencoder training to learn an internal property readout. This readout ranks and screened generated candidates by agreement with the target property, designed to help focus costly downstream evaluation on promising candidates.
\end{enumerate}

\section{EF-TALFM: a fixed-dimensional molecule-level latent with autoregressive molecular decoding}
\begingroup
\let\libertineoriginalfigure\figure
\let\libertineendoriginalfigure\endfigure
\renewenvironment{figure}[1][]{\libertineoriginalfigure[H]}{\libertineendoriginalfigure}
We first describe the canonical molecular representation used to reduce permutation and roto-translational ambiguity in \Cref{sec:mol_invariant_representation}.
We then introduce the molecule autoencoding in \Cref{sec:autoencoding}, and latent flow matching in \Cref{sec:latent_flow_matching}.
A high-level schematic of the full pipeline is shown in \Cref{fig:main_pipeline_overview}.
The full training and sampling algorithms are given in \Cref{appendix:training_sampling_algorithms}, and graph reconstruction and refinement are detailed in \Cref{appendix:molecule_reconstruction}.

\begin{figure}[ht]
    \centering
    \includegraphics[width=\linewidth]{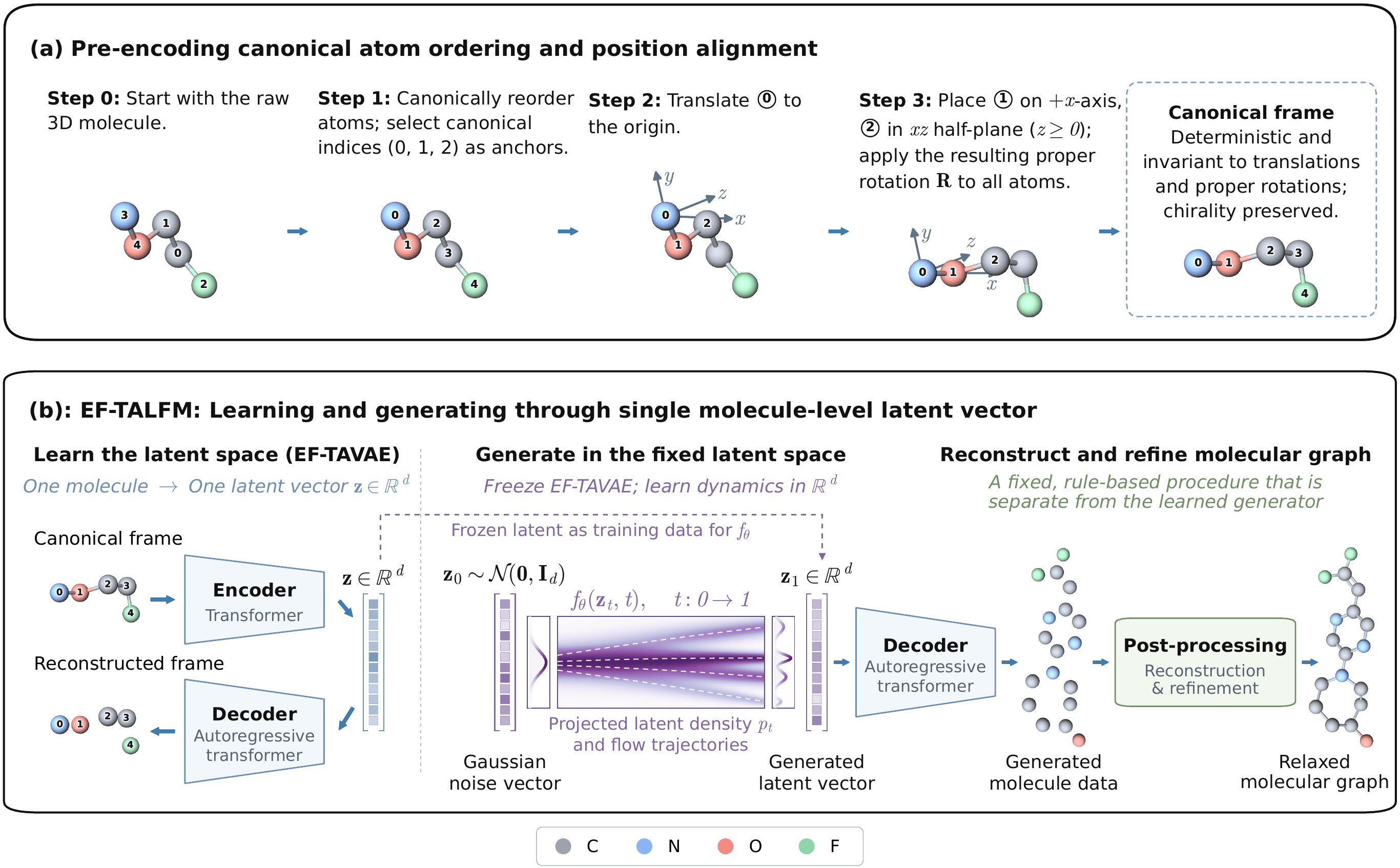}
    \caption{\small \textbf{Overview of EF-TALFM.}
    \textbf{(a) Canonical molecular representation.} Canonical atom ordering and rigid-pose alignment are illustrated with a $\mathrm{C_2H_6FNO}$ conformer from PCQM4Mv2. Hydrogen atoms are implicit and omitted from the visualization. Numbers in panel~(a) and in the canonical and reconstructed molecule views in panel~(b) denote atom-row indices. The illustrated alignment uses translations and proper rotations, preserving molecular chirality.
    \textbf{(b) Autoencoding, latent generation, and molecular reconstruction.} EF-TAVAE encodes each canonical molecule into a fixed-dimensional latent vector and reconstructs its molecular representation with an autoregressive Transformer decoder. With EF-TAVAE frozen, flow matching learns to transport Gaussian noise to the molecular latent distribution. A sampled latent vector is decoded into atom types, 3D coordinates, and atom- and molecule-level chemical attributes. A separate deterministic procedure reconstructs the molecular graph and refines its geometry.
    }
    \label{fig:main_pipeline_overview}
    \vspace{-10pt}
\end{figure}

\subsection{Canonical molecular representation}\label{sec:mol_invariant_representation}
 
Molecular data inherently exhibit two key symmetries. 
First, permutation symmetry ensures that relabeling atoms does not alter a molecule's identity. 
Second, rotation--translation symmetry implies that rotations or translations do not change the underlying structure. 
Formally, for a molecule with $n$ atoms, let $X\in\mathbb{R}^{n\times 3}$ denote the atom coordinates and $H\in\mathbb{R}^{n\times d}$ denote the corresponding per-atom features.
Let $S_n$ denote the atom-permutation group and $P_\sigma\in\{0,1\}^{n\times n}$ the matrix representing
$\sigma\in S_n$.
Let $\mathrm{SE}(3)$ denote the group of proper rigid motions. Each $g=(R,t)\in\mathrm{SE}(3)$ is specified by a rotation $R\in\mathrm{SO}(3)$ and a translation vector $t\in\mathbb{R}^3$.
Coordinates transform under both rigid motions and permutations.
Per-atom features are unchanged by rigid motions and reordered by permutations.
The joint action of $\mathrm{SE}(3)\times S_n$ on coordinates and features is given by $(g,\sigma)\cdot(X,H)=\big(P_\sigma XR^\top+\mathbf{1}_nt^\top,\;P_\sigma H\big)$, with the all-ones vector $\mathbf{1}_n\in\mathbb{R}^n$.
Representations related by this joint action describe the same molecular conformation.
Generative modeling should therefore account for these symmetries through the model or the data representation.

Molecular generators commonly handle these symmetries using equivariant networks~\citep{hoogeboom2022equivariant,xu2023geoldm,irwin2025semlaflow}, data augmentation~\citep{luo2025uae3d,joshi2025adit}, or diffusion on quotient spaces associated with the relevant symmetries~\citep{ko2026permsymdiff,xu2026quotientdiffusion}.
We instead use a deterministic \emph{canonical-alignment} procedure in preprocessing: canonical ordering fixes the atom sequence, while alignment to a standardized coordinate frame resolves translation and rotation ambiguity.
Reflections are excluded to preserve molecular chirality.

\textbf{Canonical atom ordering.}
We first establish an atom order for each molecule using \emph{canonical} SMILES. We generate an RDKit~\citep{rdkit} canonical SMILES representation, construct the corresponding canonicalized molecular graph, and match it back to the input molecule to obtain the ordering $\sigma$. 
For graphs without unresolved automorphism ties, the resulting atom ordering is deterministic and independent of the input indexing.

\textbf{Atom position alignment.} 
Once the atoms are canonically ordered, we remove translational and proper-rotational symmetries by applying a canonicalization map $c$ that places each molecule in a standardized coordinate frame.
For nondegenerate geometries, this selects a deterministic representative of the molecule's $\mathrm{SE}(3)$ orbit through rigid alignment.
The map $c$ translates the first atom to the origin and places an anchor atom on the positive $x$-axis.
It then fixes the remaining rotation about this axis by placing the first sufficiently non-collinear atom in the $x$--$z$
half-plane with $z \ge 0$.
The resulting transformation is applied to all atoms.
For a fixed canonical atom ordering, any two sets of atom coordinates differing only by a translation and proper rotation map to the same coordinates up to numerical precision, i.e., $c(g \cdot X) = c(X)$ for all $g \in \mathrm{SE}(3)$. 

By fixing the atom ordering and aligning coordinates in preprocessing (\Cref{fig:main_pipeline_overview}(a)), we shift symmetry handling
from the model architecture to the data representation. 
This provides a simple way to use standard Transformer backbones for molecule generation.
\Cref{appendix:atom_position_alignment} gives the alignment algorithm and numerical tolerances, and discusses degenerate geometries and ambiguities in canonical atom ordering.

\subsection{Molecule autoencoding} \label{sec:autoencoding}
Given the canonical molecular representation from \Cref{sec:mol_invariant_representation}, we next learn a compact latent space for 3D molecules. 
We use a variational autoencoder (VAE)~\citep{kingma2014autoencodingvariationalbayes}, referred to as EF-TAVAE, that maps each molecule to a fixed-dimensional latent vector and reconstructs it as a variable-length molecular sequence. 
Canonical ordering defines the sequence in which the decoder learns to predict atoms from the latent vector and preceding atoms.
Pose alignment standardizes the coordinates the decoder predicts.
Together, these preprocessing steps support variable-size autoregressive decoding using a standard causal Transformer~\citep{Transformer2017}.

Rather than directly predicting pairwise bond labels, the decoder predicts atom types, 3D coordinates, and a set of chemically informative atom- and molecule-level attributes.
These attributes provide electronic and topological context that helps reconstruct molecular bonds from coordinates.
The complete feature list is provided in \Cref{appendix:additional_chemical_features}.

\textbf{Molecular sequence representation.}
For a molecule with $n$ atoms, let $\mathbf{x}_i\in\mathbb{R}^3$ denote the canonicalized 3D coordinate of atom $i$, and let $\mathbf{h}_i$ denote its categorical atom-level feature vector. Atoms are ordered by a deterministic canonical ordering, and coordinates are expressed in the canonical rigid pose described in \Cref{sec:mol_invariant_representation}.
We use $k=9$ atom-level categorical fields, consisting of atom type plus eight additional atom-level chemical attributes, and $m=4$ molecule-level categorical features $\mathbf{u}$. 
Thus each molecule is represented as
\begin{equation*}
    \mathcal{M} =\big( \mathbf{u}, (\mathbf{x}_1,\mathbf{h}_1),\cdots,  (\mathbf{x}_n,\mathbf{h}_n)\big).
\end{equation*}
To enable variable-size decoding, we append a nonphysical \textit{end-of-molecule} (EOM) atom after the final atom. 
The resulting sequence has length $n+1$; the EOM atom has zero coordinates and a special end-of-sequence atom-type label, \texttt{[EOS]}, for all atom-level categorical fields. 

\textbf{Bidirectional transformer-based encoding.}
Each atom token is embedded by combining learned embeddings of the canonicalized coordinates, atom-level categorical features, and molecule-level features.
Let $\mathbf{e}^{(0)}_i\in\mathbb{R}^{d}$ denote the resulting input embedding for token $i$. 
Following BERT~\citep{bert2019Devlin}, we prepend a learnable sequence token $\mathbf{z}^{(0)}\in\mathbb{R}^{d}$ to the atom sequence $(\mathbf{z}^{(0)},\mathbf{e}^{(0)}_1,\ldots,\mathbf{e}^{(0)}_{n+1})$.
A bidirectional Transformer encoder $f_{\bm \phi}$ maps this sequence to contextualized embeddings. 
The output corresponding to the sequence token is projected to the parameters of a diagonal Gaussian posterior, $q_{\bm\phi}(\mathbf{z}\mid \mathcal{M})=\mathcal{N}(\mathbf{z}_\mu,\operatorname{diag}(\mathbf{z}_\sigma^2))$, $\mathbf{z}_\mu,\mathbf{z}_\sigma\in\mathbb{R}^{d}$.
We sample latent codes using the reparameterization trick \citep{kingma2014autoencodingvariationalbayes}, $\mathbf{z}=\mathbf{z}_\mu+\mathbf{z}_\sigma \odot \boldsymbol{\epsilon}$, $\boldsymbol{\epsilon}\sim\mathcal{N}(\mathbf{0},\mathbf{I}).$

\textbf{Masked attention transformer-based autoregressive decoding.}
The decoder factorizes molecule reconstruction into global and sequential components. 
First, a set of MLP heads predicts the molecule-level features $\hat{\mathbf{u}}$ from $\mathbf{z}$. 
Second, a causal Transformer decoder $f_{\bm \psi}$ generates atom tokens autoregressively. 
At step $i$, the decoder conditions on the latent code $\mathbf z$ and the previously generated tokens $\mathbf{x}_{<i},\mathbf{h}_{<i}$ to predict the next coordinate and atom-level attributes:
\begin{align}
    p_{\bm\psi}(\mathcal{M}\mid \mathbf{z})
    =
    p_{\bm\psi}(\mathbf{u}\mid \mathbf{z})
    \prod_{i=1}^{n+1}
    p_{\bm\psi}(\mathbf{x}_i,\mathbf{h}_i
    \mid
    \mathbf{z},\mathbf{x}_{<i},\mathbf{h}_{<i}).\label{eq:prob_molecule_given_z}
\end{align}
In particular, at rollout step $i$, the latent code and the previously generated atoms are embedded as $(\bm\nu_0^{(0)},\ldots,\bm\nu_{i-1}^{(0)})$. The Transformer decoder $f_{\bm \psi}$ contextualizes this prefix, and its last output row, $\bm\nu_{i-1}^{(L)}$, is passed to the coordinate and atom-feature readout heads to parameterize the distribution of $(\mathbf x_i,\mathbf h_i)$. For $i=n+1$, only the categorical EOM labels are modeled; $\mathbf x_{n+1}=\mathbf 0$ is a placeholder omitted from the coordinate likelihood. 
Generation stops when the decoder outputs the \texttt{[EOS]} atom-type label.
The number of atoms is therefore determined during generation rather than specified in advance.

\textbf{Property-supervised latent organization and internal property readout.}
We optionally add property supervision to molecular autoencoding. This encourages both the latent vector and the decoder's atom
representations to encode information relevant to the target property.
To predict the target property $\mathbf y\in\mathbb R^c$, we insert a learned property token $\bm\nu_{\texttt{[PROP]}}^{(0)}$ after the latent token in the decoder input.
A readout head maps its final representation $\bm\nu_{\texttt{[PROP]}}^{(L)}$ to the property prediction.

We use two complementary attention masks for property prediction.
In the molecule-conditioned route, the property token attends to atom representations to predict $\hat{\mathbf y}_{\mathrm{mol}}$, with direct attention to the latent token masked.
In the latent-only route, attention to atom tokens is masked and the prediction $\hat{\mathbf y}_{\mathrm{lat}}$ depends only on $\mathbf z$.
Under both masks, atom tokens cannot attend to the property token, preserving the autoregressive factorization in~\Cref{eq:prob_molecule_given_z}.

We use the molecule-conditioned readout for post-generation filtering.
Each generated candidate $\mathcal M^\prime$ is re-encoded and scored by the readout error relative to the target property $\mathbf y_{\mathrm{target}}$, 
\begin{align}
    e_{\mathrm{sur}}=\lVert\hat{\mathbf y}_{\mathrm{mol}}(\mathcal M^\prime)-\mathbf y_{\mathrm{target}}\rVert.
    \label{eq:readout_error}
\end{align}
For each target, we retain a chosen fraction of candidates with the lowest $e_{\mathrm{sur}}$.

\textbf{Training objective.}
The base reconstruction loss combines cross-entropy losses for categorical features with a Smooth-L1 loss for coordinates. 
We minimize
\begin{equation}
    \begin{aligned}
    \mathcal{L}_{\mathrm{rec}}
    &=
    \lambda_{\mathrm{mol}}
    \,\mathrm{CE}(\mathbf{u},\hat{\mathbf{u}})
    +
    \sum_{i=1}^{n+1}
    \lambda_{\mathrm{atom}}
    \,\mathrm{CE}(\mathbf{h}_i,\hat{\mathbf{h}}_i)
    +
    \sum_{i=1}^{n}
    \lambda_{\mathrm{coord}}
    \,\mathrm{SmoothL1}(\mathbf{x}_i,\hat{\mathbf{x}}_i),
    \end{aligned} \label{eq:recon_loss}
\end{equation}
where the coordinate term excludes the \texttt{[EOS]} token and the loss weights are chosen to balance the coordinate and categorical terms. 

When property supervision is enabled, we add a property-prediction loss with weight $\lambda_{\mathrm{prop}}$ to the base reconstruction objective in \Cref{eq:recon_loss},
\begin{align}
    \mathcal L_{\mathrm{prop}}
    &=
    \tau_{\mathrm{mol}}\,
    \mathrm{MSE}\left(
        \mathbf y,\hat{\mathbf y}_{\mathrm{mol}}
    \right)
    +
    \left(1-\tau_{\mathrm{mol}}\right)\,
    \mathrm{MSE}\left(
        \mathbf y,\hat{\mathbf y}_{\mathrm{lat}}
    \right).
    \label{eq:property_reconstruction_loss}
\end{align}
Here $\tau_{\mathrm{mol}}\in[0,1]$ is the probability of selecting the molecule-conditioned route.
 
The full VAE objective is 
\begin{align}
    \mathcal{L}_{\mathrm{VAE}}
    =
    \mathcal{L}_{\mathrm{rec}}
    +
    \lambda_{\mathrm{prop}}\,\mathcal L_{\mathrm{prop}}
    +
    \beta_{\mathrm{KL}}\,
    D_{\mathrm{KL}}
    \left(
    q_{\bm \phi}(\mathbf{z}\mid\mathcal{M})
    \,\|\, 
    \mathcal{N}(\mathbf{0},\mathbf{I})
    \right)+
    \lambda_{\mathrm{var}}\,\mathcal{L}_{\mathrm{var}}
    +
    \lambda_{\mathrm{cov}}\,\mathcal{L}_{\mathrm{cov}},\label{eq:training_objective}
\end{align}
where the property loss is the weighted average of the two routes, each categorical cross-entropy is summed over its feature heads, $\mathcal{L}_{\mathrm{var}}$ softly enforces a minimum empirical variance floor for each dimension of the posterior mean $\mathbf z_\mu$, and $\mathcal{L}_{\mathrm{cov}}$ penalizes off-diagonal covariance between dimensions of $\mathbf z_\mu$;
see \Cref{appendix:latent_regularization} for more details. We set $\lambda_{\mathrm{prop}}=0$ when property supervision is disabled.

During training, we use teacher forcing for the autoregressive decoder: when predicting token $i$, the decoder receives the ground-truth prefix $(\mathbf{x}_{<i},\mathbf{h}_{<i})$ rather than its own previous predictions. 
At inference time, the model decodes autoregressively, feeding each predicted token back into the decoder until \texttt{[EOS]} is produced.

\subsection{Latent flow matching} \label{sec:latent_flow_matching}
After training the molecule autoencoder, we freeze its parameters and train a continuous flow model~\citep{lipman2023flowmatching} in the learned latent space. For each molecule $\mathcal M$, we use the deterministic posterior mean $ \mathbf z_\mu \in\mathbb R^d$ as its latent representation (see further discussion in \Cref{appendix:learning_and_sampling_from_posterior_mean}). 
This gives an empirical target distribution over latent molecular representations, denoted $p_{\mathrm{data}}(\mathbf z_\mu)$. 
The flow model learns the marginal latent distribution $p_{\mathrm{data}}(\mathbf z_\mu)$ for unconditional generation and the conditional distribution $p_{\mathrm{data}}(\mathbf z_\mu \mid \mathbf y)$ for property-conditioned generation.

Flow matching learns a time-dependent vector field that transports a standard Gaussian base distribution $p_\varepsilon=p_0=\mathcal N(\mathbf 0,\mathbf I)$ to the target latent distribution.
For a data latent $\mathbf z_\mu\sim p_{\mathrm{data}}(\cdot)$ and a noise sample $\mathbf z_{\varepsilon}\sim p_{\varepsilon}$, we use the linear interpolation path $\mathbf z_t = (1-t)\mathbf z_{\varepsilon} + t\mathbf z_\mu$, $t\sim \mathrm{Unif}(0,1)$, with corresponding target velocity $\mathbf u_t = \tfrac{\mathrm d}{\mathrm dt}\mathbf z_t = \mathbf z_\mu-\mathbf z_{\varepsilon}$. We parametrize the velocity field by a neural network $f_\theta(\mathbf z_t,t,\mathbf y)$ and minimize the flow-matching objective
\begin{align*}
    \mathcal L_{\mathrm{FM}}(\theta)
    \;=\;
    \mathbb E_{\substack{(\mathbf z_\mu,\mathbf y)\sim p_{\mathrm{data}}(\cdot,\cdot)\\
        \mathbf z_\varepsilon\sim p_\varepsilon,\ t\sim \mathrm{Unif}(0,1)
    }}
    \left[
    \left\|
        f_\theta(\mathbf z_t,t, \mathbf y)
        -
        (\mathbf z_\mu-\mathbf z_{\varepsilon})
    \right\|_2^2
    \right].
\end{align*}
For unconditional generation, we omit $\mathbf y$ and train $f_\theta (\mathbf z_t, t)$.
At inference time, we sample $\mathbf z_\varepsilon\sim p_\varepsilon$ and solve the ordinary differential equation $\tfrac{\mathrm d}{\mathrm dt}\mathbf z_t =f_\theta(\mathbf z_t,t,\mathbf y)$ (omit $\mathbf y$ for unconditional generation) with $\mathbf z_{t=0}=\mathbf z_\varepsilon$ from \(t=0\) to \(t=1\).
The terminal state $\mathbf z_{t=1}$ is then decoded by the autoregressive molecule decoder. 
 
\label{sec:method}
\endgroup
\FloatBarrier

\section{Empirical studies}\label{sec:experiments}

We organize the empirical study around two questions.
First, can a fixed-dimensional molecule-level latent support valid, novel, and distributionally faithful generation when molecular size is determined by the decoder?
Section~\ref{sec:unconditional_generation} addresses this question by comparing EF-TALFM with baseline methods in generation quality and computational cost, then examining distributional fidelity and the contribution of postprocessing.
Second, can the same latent design support reliable property-conditioned generation?
Section~\ref{sec:property_conditioning} conditions on target HOMO--LUMO gaps and uses the model's internal property readout to rank generated candidates, measuring DFT-verified hit enrichment, candidate retention, and the novelty and diversity of the resulting hits.

\subsection{Unconditional molecule generation}\label{sec:unconditional_generation}
After describing the experimental setup in \Cref{sec:results_setup}, we compare EF-TALFM with baseline methods in \Cref{sec:results_performance_comparison} and evaluate distributional fidelity and the effects of postprocessing in \Cref{sec:results_generation_analysis}. Our results show that
(i) EF-TALFM achieves the highest validated novel yield among the 10,000 molecules generated compared to the evaluated baselines;
(ii) EF-TALFM reduces end-to-end training and sampling time while delivering the highest PoseBusters-verified novel sampling throughput and approximate training exposure throughput; and
(iii) EF-TALFM reproduces training-set distributions of molecule size and global chemical attributes, while geometry refinement improves local geometric validity without materially changing molecular identity or population-level chemical statistics.

\subsubsection{Experimental setup}\label{sec:results_setup}
We compare EF-TALFM\footnote{For unconditional generation, property supervision is disabled in EF-TAVAE.} with two bond-aware baselines: UAE-3D~\citep{luo2025uae3d}, which uses latent diffusion over variable-length atom-wise representations, and FlowMol~\citep{dunn2024flowmol}, which uses equivariant flow matching directly in 3D molecule space.
These complement EF-TALFM's fixed-dimensional molecule-level latent and chemically enriched decoding; the rationale for baseline selection and preprocessing details are given in \Cref{appendix:baseline_selection,appendix:proprocessing_pcqm4mv2}.

All three methods are trained for heavy-atom generation on PCQM4Mv2~\citep{hu2021ogblsc}, using a 90\%/10\% train/validation split after preprocessing and the same H100 hardware setting.
Each method generates $10{,}000$ samples, which undergo the same graph reconstruction and geometry refinement procedures. For UAE-3D and FlowMol, their predicted bond types are used for reconstruction.
The primary evaluation metric is the \emph{PoseBusters-verified novel yield}: the fraction of all generated samples that are unique, absent from the training set, sanitized, and pass the PoseBusters checks used here~\citep{Buttenschoen2024posebuster}.
This joint metric combines uniqueness, novelty, and chemical and geometric validity, providing a more comprehensive assessment of generation quality and therefore a more informative measure for molecular discovery than any criterion considered separately.
Training budgets and sampling settings are detailed in \Cref{appendix:unconditional_comparison_protocol}, and evaluation metrics are defined in \Cref{appendix:evaluation_metrics}.

\subsubsection{Performance comparison against baselines} \label{sec:results_performance_comparison}
\begin{figure}[t]
    \centering
    \includegraphics[width=1.0\linewidth]{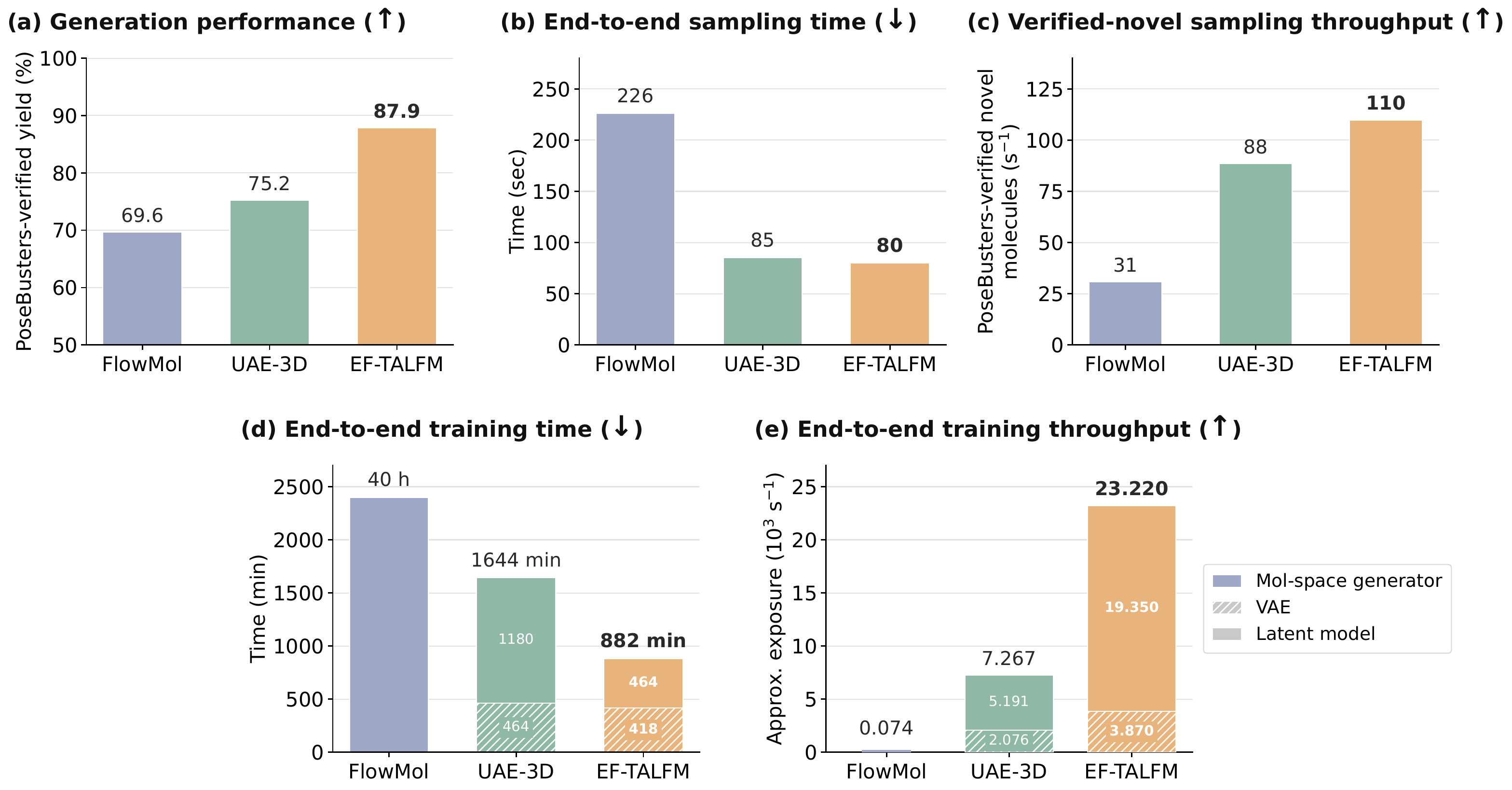}
    \caption{\small
    \textbf{Generation quality and end-to-end computational efficiency on PCQM4Mv2.}
    Each method generates $10{,}000$ samples under the same H100 hardware setting; arrows indicate the preferred direction.
    \textbf{(a)} PoseBusters-verified novel yield, normalized over all generated samples.
    \textbf{(b)} End-to-end sampling time, including generation, graph reconstruction, and geometry refinement.
    \textbf{(c)} Throughput of PoseBusters-verified novel molecules.
    \textbf{(d)} Total training time, with autoencoder and latent-model stages distinguished for the two-stage methods.
    \textbf{(e)} Approximate training-exposure throughput. EF-TALFM and UAE-3D count molecule examples processed across both training stages; FlowMol uses an estimated molecule-equivalent exposure derived from sampled edges. }
    \label{fig:generation_compute_efficiency_summary}
\end{figure}

\textbf{Higher verified novel yield at lower computational cost.}
EF-TALFM achieves the highest PoseBusters-verified novel yield among the evaluated methods, $87.9\%$, compared with $75.2\%$ for UAE-3D and $69.6\%$ for FlowMol (\Cref{fig:generation_compute_efficiency_summary}a).
This higher yield is achieved at lower end-to-end sampling cost (\Cref{fig:generation_compute_efficiency_summary}b,c).
EF-TALFM achieves end-to-end sampling speedups of $1.06\times$ over UAE-3D and $2.83\times$ over FlowMol.
Together with its higher yield, these speedups translate into PoseBusters-verified novel throughput of $1.24\times$ and $3.56\times$ that of the respective baselines.
Further results are provided for the individual evaluation criteria (\Cref{appendix:unconditional_individual_criteria_comparisons}), baseline checkpoints and sampling-step budgets (\Cref{appendix:unconditional_additional_baseline}), and generator-only sampling times (\Cref{appendix:unconditional_generator_sampling_times}).

Training is also faster under the reported budgets (\Cref{fig:generation_compute_efficiency_summary}d,e).
EF-TALFM reduces end-to-end training time by factors of $1.86$ relative to UAE-3D and $2.72$ relative to FlowMol.
Its training-exposure throughput is $3.20\times$ that of UAE-3D and approximately $314\times$ that of FlowMol, with the latter comparison based on an estimated molecule-equivalent exposure count.
Compared with UAE-3D, most of the training-time reduction occurs in the latent generative stage, consistent with modeling a single molecule-level vector rather than a variable-size atom-wise latent array.
The training budgets, stage-level training times and exposure calculations are provided in \Cref{appendix:unconditional_training_cost}.

\subsubsection{Distributional fidelity and postprocessing ablation} \label{sec:results_generation_analysis}
\begin{figure}[htbp]
    \centering
    \includegraphics[width=\linewidth]{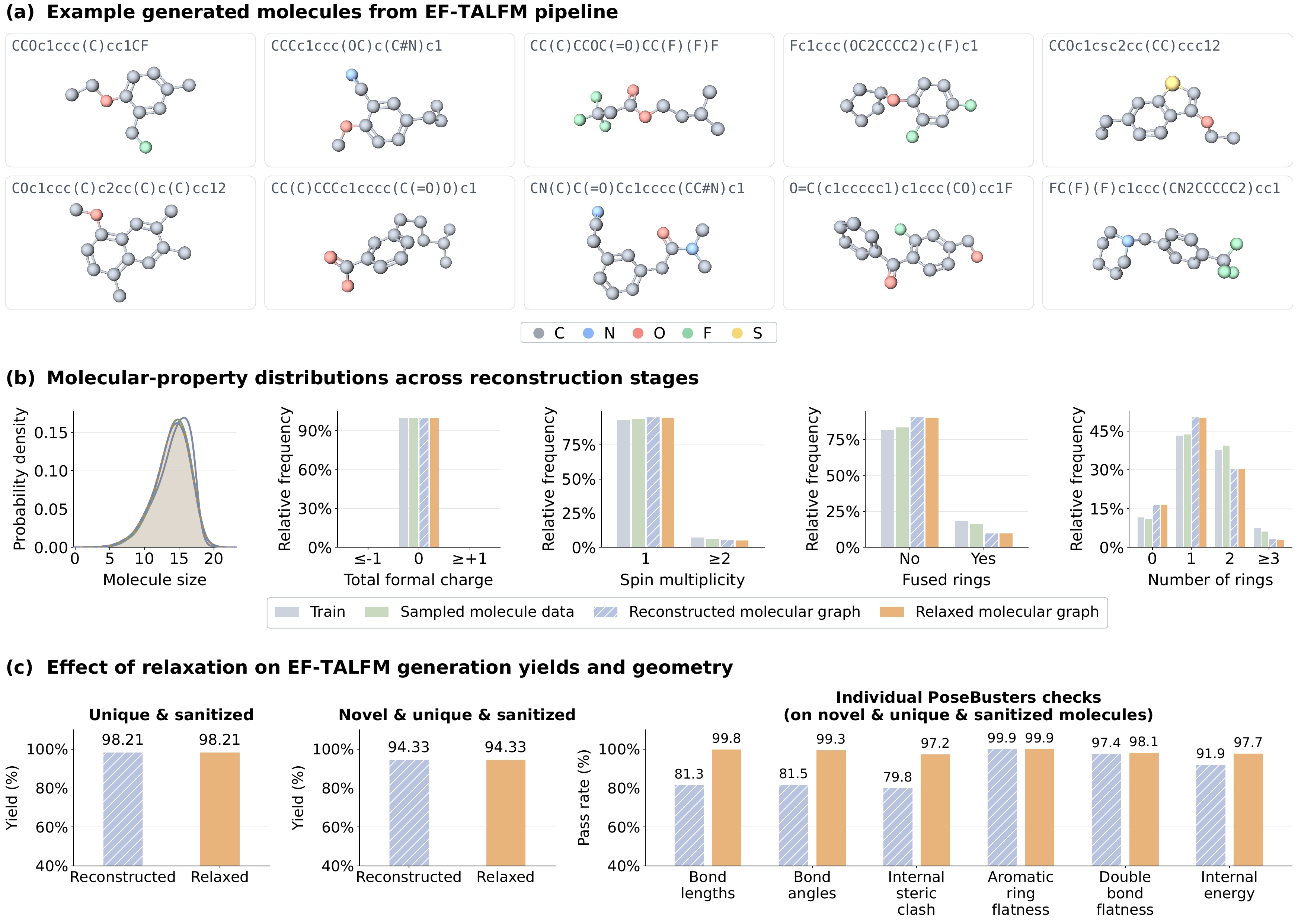}
    \caption{\small \textbf{Example molecules generated from the EF-TALFM pipeline, learned chemical trends, and geometry refinement.}
    \textbf{(a)} Ten novel generated molecules, each labeled with its canonical SMILES. 
    \textbf{(b)} Distributions of molecule size and chemical attributes across sampled latent decoding, graph reconstruction, geometry refinement. 
    \textbf{(c)} Generation yield and geometric validity before and after geometry refinement. Generation yields are normalized over all generated samples, and individual PoseBusters checks are evaluated within the novel, unique, sanitized subset.
    }
    \label{fig:generation_distribution_and_ablation_unconditional}
\end{figure}

\Cref{fig:generation_distribution_and_ablation_unconditional}a shows ten example molecules generated by EF-TALFM pipeline, spanning different ring systems and functional groups.
We next assess how well EF-TALFM generates molecules that reflect the chemical-property distributions of the training data and how postprocessing changes the generated distributions.
We compare $10{,}000$ generated samples with the training data across decoding, graph reconstruction, and geometry refinement. 
This comparison covers molecule size and the four molecule-level chemical attributes predicted by the decoder. 

\textbf{Postprocessing broadly retains the learned chemical-property profiles.}
The main chemical-property trends are captured during generation and largely preserved through postprocessing (\Cref{fig:generation_distribution_and_ablation_unconditional}b).
The EF-TALFM generation pipeline proceeds through decoding sampled latents, graph reconstruction, and geometry refinement.
Among these three stages, the decoded samples most closely match the training-set distributions of molecule size and chemical attributes.
Graph reconstruction introduces shifts, most visibly in ring count and fused-ring frequency, but the distributions remain broadly similar and do not change further during geometry refinement.

\textbf{Uniqueness, chemical validity, and novelty are preserved across stages, while geometry refinement improves molecular geometry.}
We then quantify the effects of geometry refinement on generation yield and geometric validity.
Geometry refinement improves molecular geometry while preserving chemical validity, uniqueness, and novelty (\Cref{fig:generation_distribution_and_ablation_unconditional}c).
The yield of unique molecules passing sanitization remains unchanged, as does the yield of those absent from the training set.
Among molecules meeting all three criteria, individual PoseBusters pass rates increase by up to $18.5$ percentage points.
The largest gains occur in bond-length, bond-angle, and internal steric-clash checks, consistent with the correction of local geometric distortions.

\subsection{Property-conditioned molecule generation} \label{sec:property_conditioning}

Having established EF-TALFM's performance in unconditional generation, we next assess its ability to generate molecules with desired properties.

Using the protocol in \Cref{sec:condition_experimental_setup}, we evaluate DFT-verified property targeting, novelty, and structural diversity (\Cref{sec:condition_generation_hits_evaluation}).
We show that internal-readout selection
(i) narrows the distributions of DFT-computed properties around the requested targets;
(ii) increases the proportion of verified hits while retaining most available hits; and
(iii) largely preserves novelty and structural diversity among the verified hits.

\subsubsection{Experimental setup and evaluation protocol} \label{sec:condition_experimental_setup}

On PCQM4Mv2, we condition generation on ten target HOMO-LUMO gaps between $4.1$ and $7.8\,\mathrm{eV}$, spanning the 10th--90th percentiles of the training-set distribution. 
Most molecules in the dataset have no explicitly assigned radical electrons, and we focus on this subset.
We use an all-atom representation with explicit hydrogen coordinates to provide complete geometries for single-point DFT calculations. 
See \Cref{appendix:conditional_data_targets} for dataset construction and preprocessing details.
Property supervision is enabled during EF-TAVAE training, and the latent flow is conditioned on the target gap.

We generate $10{,}000$ samples at each of the ten HOMO--LUMO gap targets, giving $100{,}000$ molecular structures in total.
After graph reconstruction and geometry refinement, we focus our property evaluation on $77{,}724$ strictly sanitized molecules that pass the PoseBusters geometric checks and contain no explicit radical electrons as in the training set ($\mathcal{M}_{\mathrm{gen}}$).
For each target property value $y_{\mathrm{target}}$, we retain the $30\%$ of screened candidates whose internal-readout property estimates are closest to that target value, leaving $23{,}315$ selected molecules ($\widehat{\mathcal{M}}_{\mathrm{gen}}$) after pooling across targets.
We evaluate both populations by measuring the absolute targeting error, $\varepsilon_{\mathrm{true}}=\vert y_{\mathrm{true}}-y_{\mathrm{target}}\vert$, where $y_{\mathrm{true}}$ is computed using single-point DFT calculations at the \texttt{B3LYP/6-31G(d)} level as implemented in \texttt{Psi4}~\citep{psi4_2020}.
We quantify targeting accuracy using hit rates at $0.1$ and $0.2\,\mathrm{eV}$ thresholds, with the size of each population as its denominator.
Detailed screening counts and DFT evaluation totals are provided in \Cref{appendix:conditional_population_accounting}.

\begin{figure}[t]
    \centering
    \includegraphics[width=1.0\linewidth]{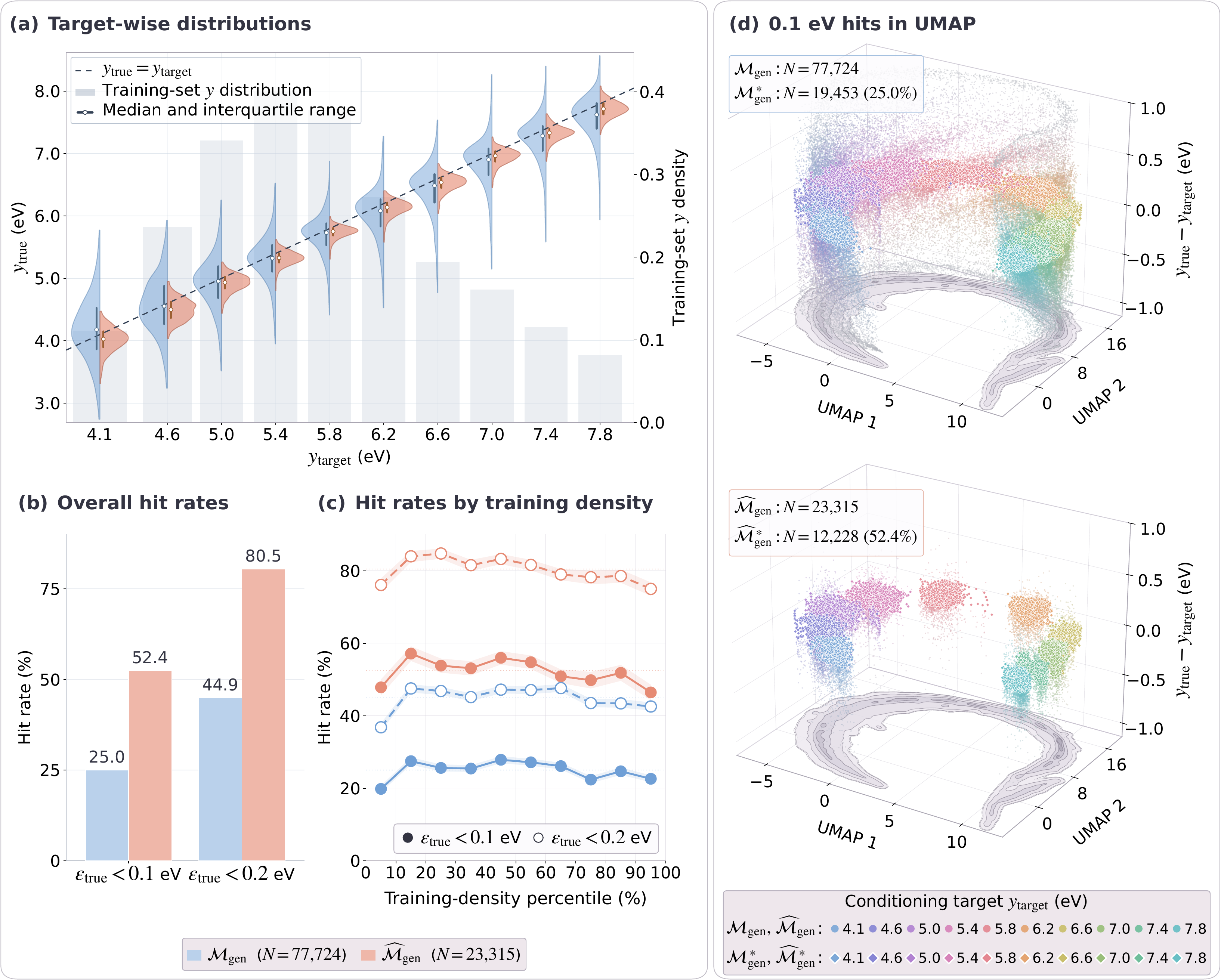}
    \caption{\small
    \textbf{Internal-readout selection improves property targeting.}
    Results compare $77{,}724$ screened molecules ($\mathcal{M}_{\mathrm{gen}}$) with $23{,}315$ selected molecules ($\widehat{\mathcal{M}}_{\mathrm{gen}}$), pooled across ten conditioning targets.
    \textbf{(a)} Distributions of DFT-computed HOMO--LUMO gaps ($y_{\mathrm{true}}$) at each target ($y_{\mathrm{target}}$), before (blue, left halves) and after (orange, right halves) selection. Circles and thick bars mark medians and interquartile ranges; the dashed line indicates exact target agreement, and gray bars show the training-set gap distribution.
    \textbf{(b)} Hit rates for absolute targeting errors $\varepsilon_{\mathrm{true}}$ below $0.1$ and $0.2\,\mathrm{eV}$, normalized by the size of each population.
    \textbf{(c)} Hit rates across ten bins of local training-density percentile in a shared UMAP projection of the learned property representation, pooled across targets. Filled and open circles denote the $0.1$ and $0.2\,\mathrm{eV}$ thresholds, respectively; shaded bands show $95\%$ Wilson confidence intervals, and dotted lines indicate overall rates.
    \textbf{(d)} Screened (top) and selected (bottom) populations in the same UMAP projection, with signed targeting error $y_{\mathrm{true}}-y_{\mathrm{target}}$ shown vertically and clipped at $\pm1.0\,\mathrm{eV}$. White-outlined diamonds denote the corresponding $0.1\,\mathrm{eV}$ hit populations, $\mathcal{M}_{\mathrm{gen}}^\ast$ and $\widehat{\mathcal{M}}_{\mathrm{gen}}^\ast$; circles denote other candidates. 
    Colors identify conditioning targets and fade toward gray with increasing absolute error. 
    Floor shading shows full-training-set density, with contours enclosing the highest-density regions containing $\{1,10,50,90,99,99.5\}\%$ of training molecules.}
    \label{fig:internal_selection_and_umap}
\end{figure}

\subsubsection{Reference-verified targeting and structural diversity}\label{sec:condition_generation_hits_evaluation}

\textbf{Internal-readout selection narrows property distributions at each target and increases hit rates.}
Selection narrows the distributions of DFT-evaluated HOMO--LUMO gaps at all ten targets and increases the fraction of accurate candidates (\Cref{fig:internal_selection_and_umap}a,b).
At the $0.1\,\mathrm{eV}$ threshold, the screened and selected hit subsets contain $19{,}453$ molecules ($\mathcal{M}_{\mathrm{gen}}^\ast$) and $12{,}228$ molecules ($\widehat{\mathcal{M}}_{\mathrm{gen}}^\ast$), respectively, corresponding to hit rates of $25.0\%$ and $52.4\%$.
At $0.2\,\mathrm{eV}$, the hit rate rises from $44.9\%$ to $80.5\%$.
The higher hit rates result from the retention of most hits alongside the removal of most off-target candidates.
At the $0.1\,\mathrm{eV}$ threshold, selection retains $62.9\%$ of the available hits but only $19.0\%$ of off-target candidates (\Cref{fig:internal_selection_and_umap}d).
The trade-off between candidate coverage and DFT-verified hit rate at different retention thresholds is detailed in \Cref{appendix:selection_at_different_retention_thresholds}.

To assess how these gains vary with training density, we compare hit rates across ten density-percentile bins.
We estimate local training density from the full training set in a shared two-dimensional uniform manifold approximation and projection (UMAP)~\citep{mcinnes2018umap} of the learned property representations.
\Cref{fig:internal_selection_and_umap}(c) shows that selection increases hit rates in every density bin at both $0.1$ and $0.2\,\mathrm{eV}$ tolerances, demonstrating that its benefit is not confined to the densest regions of this projection, and extends across sparsely and densely populated regions of the projected training distribution.
See \Cref{appendix:conditional_density_analysis} for details of the density estimation and binning procedure.
A complementary principal component analysis yields consistent findings; see \Cref{appendix:pca_visualization} for details.

\begin{table}[t]
    \centering
    \caption{\small
    \textbf{Internal-readout selection preserves 97.4\% exact-match novelty among unique verified hits, with only a slight reduction in scaffold novelty and internal diversity.}
    Screened hits, $\mathcal{M}_{\mathrm{gen}}^\ast$, and selected hits, $\widehat{\mathcal{M}}_{\mathrm{gen}}^\ast$, are the subsets of $\mathcal{M}_{\mathrm{gen}}$ and $\widehat{\mathcal{M}}_{\mathrm{gen}}$, respectively, whose DFT-computed HOMO-LUMO gaps lie within $0.1\,\mathrm{eV}$ of their targets.
    In the upper block, $N$ counts hits before deduplication, and U. is the percentage of unique molecules. U.\&\,N. is the percentage of unique molecules absent from the property-matched training reference by canonical-SMILES matching; each reference contains training molecules within $0.1\,\mathrm{eV}$ of the same target.
    Maximum Tanimoto is each unique molecule's highest ECFP4 fingerprint similarity to this reference, summarized by the median (interquartile range, IQR) and percentages below $0.4$ and $0.8$.
    Scaffold novelty is the percentage of distinct nonempty Bemis--Murcko scaffolds absent from the reference.
    In the lower block, $\widetilde{\mathcal{M}}_{\mathrm{train}}^\ast$ and $\widehat{\mathcal{M}}_{\mathrm{train}}^\ast$ denote training controls matched within each target to the numbers of unique screened and selected hits, respectively.
    $N_{\mathrm{uniq}}$ is the number of unique molecules counted within each target and summed across targets. Mean similarity is calculated over unordered within-target molecular pairs; internal diversity is one minus this mean.
    Training-control values are means $\pm$ sample standard deviations over 20 resamples. Counts are summed across targets; molecular comparisons are made within targets before pooling.
    } \label{tbl:generated_novelty_internal_diversity}

    \small
    \setlength{\tabcolsep}{5.2pt}
    \renewcommand{\arraystretch}{1.14}

    {
    \sisetup{
        detect-weight=true,
        group-separator={,},
        group-digits=integer,
        group-minimum-digits=4
    }

    \begin{tabular}{
        @{}l
        S[table-format=5.0, group-separator={,}, group-minimum-digits=4]
        S[table-format=2.1]
        S[table-format=2.1]
        c
        S[table-format=2.1]
        S[table-format=2.1]
        S[table-format=2.1]
        @{}
    }
    \toprule
    &
    \multicolumn{3}{c}{}
    &
    \multicolumn{3}{c}{Maximum Tanimoto}
    &
    \multicolumn{1}{c}{}
    \\
    \cmidrule(lr){2-4}
    \cmidrule(lr){5-7}

    \multicolumn{1}{c}{Cohort}
    & \multicolumn{1}{c}{$N$}
    & \multicolumn{1}{c}{U. (\%)}
    & \multicolumn{1}{c}{U.\&\,N. (\%)}
    & \multicolumn{1}{c}{Median (IQR)}
    & \multicolumn{1}{c}{$<0.4$ (\%)}
    & \multicolumn{1}{c}{$<0.8$ (\%)}
    & \multicolumn{1}{c}{Scaffold novelty (\%)}
    \\
    \midrule
    $\mathcal{M}_{\mathrm{gen}}^\ast$
    & 19453
    & 99.4
    & 97.7
    & 0.556 (0.173)
    & 8.9
    & 94.6
    & 37.6
    \\
    \midrule
    $\widehat{\mathcal{M}}_{\mathrm{gen}}^\ast$
    & 12228
    & 99.4
    & 97.4
    & 0.568 (0.171)
    & 6.8
    & 94.1
    & 33.5
    \\
    \bottomrule
    \end{tabular}

    \vspace{0.9em}

    \begin{tabular}{
        @{}l
        r
        r
        S[table-format=1.4(2), separate-uncertainty=true]
        S[table-format=1.4(2), separate-uncertainty=true]
        @{}
    }
    \toprule
    & \multicolumn{1}{c}{$N_{\mathrm{uniq}}$}
    & \multicolumn{1}{c}{Within-target unordered pairs}
    & \multicolumn{1}{c}{Mean similarity}
    & \multicolumn{1}{c}{Internal diversity}
    \\
    \midrule
    $\widetilde{\mathcal{M}}_{\mathrm{train}}^\ast$
    & \num{19338}
    & \num{19725045}
    & 0.0843(2)
    & 0.9157(2)
    \\ 
    $\mathcal{M}_{\mathrm{gen}}^\ast$
    & \num{19338}
    & \num{19725045}
    & 0.1161
    & 0.8839
    \\
    \midrule
    $\widehat{\mathcal{M}}_{\mathrm{train}}^\ast$
    & \num{12154}
    & \num{7732833}
    & 0.0844(3)
    & 0.9156(3)
    \\ 
    $\widehat{\mathcal{M}}_{\mathrm{gen}}^\ast$
    & \num{12154}
    & \num{7732833}
    & 0.1224
    & 0.8776
    \\
    \bottomrule
    \end{tabular}
    }
\end{table}

\textbf{Readout-selected hits remain structurally novel and diverse.}
Selection largely preserves the novelty and structural diversity of DFT-verified hits (\Cref{tbl:generated_novelty_internal_diversity}).
Uniqueness remains $99.4\%$ after selection, and $97.4\%$ of unique selected hits are absent from their corresponding property-matched training subsets under exact canonical-SMILES matching, compared with $97.7\%$ before selection.
Here, each reference subset contains training molecules whose gaps lie within $0.1\,\mathrm{eV}$ of the same target.
Nearest-training fingerprint similarity increases only slightly, while scaffold novelty decreases from $37.6\%$ to $33.5\%$.
Internal diversity decreases from $0.8839$ to $0.8776$, a small change relative to the full hit population.
Analyses at individual conditioning targets lead to the same conclusion (\Cref{appendix:condition_ablation_study}).
Detailed definitions for the novelty and diversity metrics are given in \Cref{appendix:conditional_novelty_and_diversity}.

\FloatBarrier

\section{Discussion}\label{sec:discussion}
EF-TALFM generates variable-size 3D molecules by sampling a single fixed-dimensional molecule-level latent vector, with molecular size and atom-wise structure determined only during autoregressive decoding.
Canonicalization allows EF-TALFM to use standard Transformer backbones without equivariant layers or data augmentation.
Predicted chemical attributes guide bond reconstruction without requiring a learned dense bond-prediction head.

On PCQM4Mv2, compared to baselines, EF-TALFM gives the strongest overall generation performance with shorter end-to-end training and sampling times for unconditional generation.
For property-conditioned generation, EF-TALFM's internal-readout selection approximately doubles the fraction of structurally screened molecules within the target property tolerances, while largely preserving novelty and structural diversity.

These results suggest that the choice of molecular representation can simplify the architecture of 3D generative models.
Future theoretical work could characterize the consistency of canonical atom ordering for symmetric molecules and the stability of alignment near degenerate geometries.
It is also worth exploring learned bond refinement to improve chemical consistency while preserving the fixed-dimensional latent design.

EF-TALFM provides a practical approach to discovering molecules with desired properties by combining molecular generation with internal candidate selection.
Incorporating uncertainty estimates for the internal readout and feedback from calculations or experiments would enable an active-learning cycle for exploring properties beyond the initial training distribution.
Adapting its representations and autoregressive decoding from latent vectors could extend this approach to a broader range of tasks, including organic ligand and organocatalyst design.

\section*{Code availability}
The code implementing EF-TALFM and reproducing the experimental results is available in the github repository \href{https://github.com/weichiyao/EF-TALFM}{https://github.com/weichiyao/EF-TALFM}.

\section*{Acknowledgments}

This work was supported in part by funding from the Office of Naval Research under grant N00014-23-1-2590, the National Science Foundation under grant No. 2310831, No. 2428059, No. 2435696, No. 2440954, a Michigan Institute for Data Science Propelling Original Data Science (PODS) grant, Two Sigma Investments LP, and  LG Management Development Institute AI Research. This work used computing resources from the Advanced Cyberinfrastructure Coordination Ecosystem: Services \& Support (ACCESS) program, which is supported by the U.S. National Science Foundation.

\clearpage
\bibliographystyle{unsrtnat}
\bibliography{reference_cleaned}
\clearpage

\appendix
\crefalias{section}{appendix}
\crefalias{subsection}{appendix}
\crefalias{subsubsection}{appendix}

\begingroup
\let\small\scriptsize
\AtBeginEnvironment{tabular}{\fontsize{7}{8.5}\selectfont}
\section[Unconditional generation: protocols and supporting results]{Unconditional generation: protocols and\\supporting results}\label{appendix:unconditional}
\subsection{Experimental setup and evaluation}\label{appendix:unconditional_experimental_setup}
\subsubsection{Baseline selection}  \label{appendix:baseline_selection}
We select two strong, complementary bond-aware baselines, UAE-3D~\citep{luo2025uae3d} and FlowMol~\citep{dunn2024flowmol}.
Across their respective evaluations, both report state-of-the-art results for \emph{de novo} 3D molecular generation, collectively outperforming a broad range of prior 3D generators~\citep{hoogeboom2022equivariant,xu2023geoldm,huang2023jodo,vignac2023midi,hua2024mudiff,le2024eqgatdiff,joshi2025adit,irwin2025semlaflow}.
They cover two complementary generative settings: UAE-3D learns a variable-length atom-wise latent representation that is later modeled by a latent diffusion model,
whereas FlowMol uses equivariant flow matching to directly generate molecular structures, including bond types.

We do not include other methods that generate only atom types and 3D coordinates~\citep{hoogeboom2022equivariant,xu2023geoldm,joshi2025adit,cheng2025quetzal}, because geometry-based graph recovery can be insufficient for accurately reconstructing chemically annotated molecular states, particularly for larger or heavy-atom-rich molecules where bond connectivity, bond orders, aromaticity, and formal charges may be ambiguous from coordinates alone; see \Cref{tbl:reconstruction_comparison_BOND_OR_NOT}.

ADiT \citep{joshi2025adit} is a closely related work that uses a scalable latent Transformer framework without explicit molecular graph generation.
However, for non-periodic molecules, ADiT decodes only atom types and 3D coordinates and relies on RDKit-based post-processing to infer the molecular graph.
Such geometry-based graph recovery can be insufficient for accurately reconstructing chemically annotated molecular states, particularly for larger or heavy-atom-rich molecules where bond connectivity, bond orders, aromaticity, and formal charges may be ambiguous from coordinates alone; see \Cref{tbl:reconstruction_comparison_BOND_OR_NOT}.
We therefore treat ADiT as an important architectural reference rather than a main quantitative baseline.

\begin{table}[H] 
    \captionof{table}{
    {\textbf{Generated bond types improve reconstruction from atom types and 3D coordinates.} Including generated bond types raises UAE-3D exact SMILES reconstruction from 76.8\% to 99.5\% while leaving sanitization nearly unchanged.
    }}
    \label{tbl:reconstruction_comparison_BOND_OR_NOT}
    \centering
    \begin{tabular*}{0.9\linewidth}{@{\extracolsep{\fill}}lccc@{}}  
        \toprule
            & Connectivity & Sanitization  & SMILES match \\
            \midrule 
            Atom type + coordinates + bond & 100.0\% & 99.5\% & 99.5\% \\
            Atom type + coordinates & 99.0\%& 99.2\% & 76.8\%\\
            \bottomrule
    \end{tabular*}  
\end{table}

\subsubsection{Details of PCQM4Mv2 dataset preprocessing}\label{appendix:proprocessing_pcqm4mv2}
We use PCQM4Mv2~\citep{hu2021ogblsc}, a large-scale quantum-chemistry dataset derived from PubChemQC and originally designed for HOMO--LUMO gap prediction in the OGB Large-Scale Challenge; its official training split provides DFT-optimized 3D structures for approximately 3.4M molecules.
Following common practice in 3D small-molecule generation, we model only the heavy-atom scaffold.

Although PCQM4Mv2 is commonly treated as a predominantly neutral, closed-shell small-molecule corpus, our feature extraction reveals a small number of non-singlet entries. We therefore model spin multiplicity explicitly as a molecule-level feature and restrict the training corpus to multiplicity $\leq 4$, retaining singlet through quartet states while excluding the extremely sparse high-spin tail. This filtering defines a well-supported chemical domain for generative modeling rather than declaring the excluded molecules invalid.
After preprocessing, we use a 90\%/10\% train/validation split for all three methods.

\subsubsection{Training, sampling, and comparison protocol}\label{appendix:unconditional_comparison_protocol}
The methods operate in different representation spaces and therefore do not share directly comparable notions of batch size, epoch, or training step. UAE-3D and EF-TALFM use two-stage training, in which an autoencoder is followed by an unconditional latent generative model (with property supervision disabled for EF-TAVAE), whereas FlowMol trains a direct molecular graph flow model using adaptive graph-edge batches. We therefore use a common evaluation protocol and hardware while reporting each method's native optimizer-step budget, parameter count, effective update units, and wall-clock time.

For the PCQM4Mv2 generation results reported below, we train the $30$M-parameter FlowMol model for 100K optimizer steps and evaluate its validation-selected checkpoint. For UAE-3D, we train the $0.6$M-parameter VAE for 400K steps and the $38.6$M-parameter latent generative model for 1M steps; for EF-TALFM, we train the $18.5$M-parameter TAVAE for 400K steps and the $5.4$M-parameter latent flow model for 1M steps. We adapt the published UAE-3D and FlowMol QM9 configurations to PCQM4Mv2, with minor tuning for stable training. EF-TALFM uses a fixed molecule-level latent dimension of $d=32$, whereas UAE-3D uses $d=4$ dimensions per atom; because most PCQM4Mv2 molecules contain more than 11 atoms, UAE-3D typically operates on a higher-dimensional molecule-level latent representation. All models are trained on single-node H100 GPUs. Detailed hyperparameters are provided in \Cref{appendix:hyperparameter_uae_3d,appendix:hyperparameter_flowmol,appendix:hyperparameter_talfm}.

For sampling, EF-TALFM integrates its latent flow ODE with the adaptive fifth-order Dormand--Prince solver implemented in \texttt{torchdiffeq}~\citep{torchdiffeq}, UAE-3D uses its denoising diffusion probabilistic model (DDPM) sampler with 100 denoising steps, and FlowMol uses 100 evenly spaced Euler integration steps.
We measure the time required to generate 10,000 molecules using batch size 5,000 under the same hardware setting.
EF-TALFM samples are reconstructed and refined using the procedure in~\Cref{appendix:molecule_reconstruction}; UAE-3D and FlowMol use the same downstream reconstruction and refinement stages with their predicted bond types.

End-to-end sampling times include generation, graph reconstruction, and geometry refinement.
Training times include both the autoencoder and latent generator for the two-stage methods.
Training-exposure throughput is an approximate data-volume measure rather than a FLOP count; the method-specific calculations and full timings are provided in \Cref{tbl:gradient_evals}.

\subsubsection{Evaluation metrics}\label{appendix:evaluation_metrics}
The primary metric, PoseBusters-verified novel yield, counts unique, training-set-novel molecules that pass strict sanitization and all six PoseBusters sanity checks used here~\citep{Buttenschoen2024posebuster}, divided by all generated samples.
For fragmented outputs, SMILES-based and PoseBusters metrics use the atom-count largest connected component after sanitization.
Connectivity, sanitization, unique sanitized yield and novel unique sanitized yield use all generated samples as their denominator; individual PoseBusters check rates use the novel, unique, sanitized subset.
Missing internal-energy evaluations count as failures.
We use the six checks listed below from the PoseBusters \texttt{mol} configuration.
The component metrics and thresholds used in our evaluation pipeline are:
\begin{itemize}
    \item Connectivity: \% of molecules with exactly one connected component, i.e., one fragment.
    \item Sanitization: \% of molecules with a canonical SMILES string produced by RDKit that passed the strict sanitization check.
    \item Uniqueness: \% of unique sanitized SMILES among all generated ones.
    \item Novelty: \% of novel unique sanitized SMILES against the training data set.
    \item Reasonable bond lengths: \% of molecules for which all bond lengths lie between 0.75 times the lower and 1.25 times the upper RDKit distance-geometry bounds.
    \item Reasonable bond angles: \% of molecules for which all angle-defining 1,3-atom distances lie between 0.75 times the lower and 1.25 times the upper RDKit distance-geometry bounds.
    \item Aromatic ring flatness: \% of molecules for which all atoms in 5- and 6-membered aromatic rings lie within $0.25\,\mathring{\mathrm{A}}$ of the closest shared plane.
    \item Double bond flatness: \% of molecules for which the atoms in trigonal-planar aliphatic $\mathrm{C}=\mathrm{C}$ systems and their neighbors lie within $0.25\,\mathring{\mathrm{A}}$ of the closest shared plane.
    \item Reasonable internal energy: \% of molecules for which the UFF energy ratio is at most 100 relative to the average energy of an ensemble of 50 generated reference conformations.
    \item No internal steric clash: \% of molecules for which all non--covalently bound heavy-atom pairs are separated by at least 0.7 times the distance--geometry lower bound.
\end{itemize}

\textbf{Selection of illustrated molecules.}
The ten examples in \Cref{fig:generation_distribution_and_ablation_unconditional}a are selected from chemically screened generated structures with no exact matches in either the full PCQM4Mv2 dataset or the processed molecular-graph corpus.
Identity is compared using RDKit canonical non-isomeric SMILES after removing explicit hydrogens, retaining bond orders and formal charges without tautomer normalization.
The chemical screen requires connected, neutral graphs with no explicit radical electrons, 9--22 heavy atoms drawn from C, N, O, F, S and Cl, at least one N, O or S atom, at most three five- or six-membered rings, at most five rotatable bonds, a synthetic-accessibility score of at most 3, and no BRENK, NIH or PAINS structural alerts.
Bridgehead and spiro atoms, nonaromatic $\mathrm{C}=\mathrm{C}$ and $\mathrm{C}=\mathrm{N}$ bonds, and nonaromatic bonds between N, O and S atoms are excluded.
The selected examples span different ring systems and functional groups and pass all six PoseBusters checks above; their saved coordinates and connectivity are preserved for visualization.
These screening criteria indicate plausible chemistry but do not establish experimental stability or a validated synthetic route.

\subsection{Comparisons across individual evaluation criteria}
\label{appendix:unconditional_individual_criteria_comparisons}
\Cref{tbl:pcqm4mv2_generation} shows that relative to FlowMol, EF-TALFM nearly matches the sanitization rate (98.25\% versus 98.72\%) and unique sanitized yield (98.21\% versus 98.27\%), while its connectivity is moderately lower (92.98\% versus 98.72\%).
Meanwhile, EF-TALFM outperforms UAE-3D, its closest latent-space counterpart, on every reported molecule-level generation metric.
Despite not generating explicit bond labels, EF-TALFM also remains close to both bond-aware baselines in 3D physical plausibility, passing each of the six geometric and chemical-validity checks from PoseBusters~\citep{Buttenschoen2024posebuster} at rates between 97.22\% and 99.93\% among molecules that are sanitized, unique, and absent from the training set.
\begin{table}[ht]
    \centering
    \caption{
    \textbf{EF-TALFM has the highest yield of sanitized, unique, and training-set-novel molecules, while remaining competitive on validity without generating bond labels.}
    N.\,\&\,U.\,\&\,S. counts generated candidates that are sanitized, nonduplicate and absent from the training set, before requiring the joint PoseBusters pass used in the primary verified novel yield. EF-TALFM achieves 94.33\%, versus 76.03\% for UAE-3D and 69.83\% for FlowMol.
    UAE-3D is the closest latent counterpart but uses variable-length atom-wise latents and predicts explicit bond types; FlowMol directly generates molecular graphs, including bond types. EF-TALFM instead generates no bond labels and recovers a graph deterministically from its decoded geometry and chemically enriched state.
    In (a), EF-TALFM outperforms UAE-3D on every metric and nearly matches FlowMol in sanitization and unique sanitized yield, with moderately lower connectivity.
    In (b), EF-TALFM nevertheless attains 97.22--99.93\% pass rates across the six individual PoseBusters checks, close to the bond-explicit baselines, and 93.13\% pass all checks jointly.
    Panel (a) rates are computed over 10,000 generated molecules. C. denotes the fraction of connected molecules; S. the fraction passing strict RDKit sanitization; U.\,\&\,S. the fraction of unique sanitized canonical SMILES; and N.\,\&\,U.\,\&\,S. the fraction that are sanitized, unique, and absent from the training set. Panel (b) PoseBusters rates are computed among the N.\,\&\,U.\,\&\,S. molecules: 7,603 for UAE-3D, 6,983 for FlowMol, and 9,433 for EF-TALFM. Missing internal-energy evaluations (45 for UAE-3D, 12 for FlowMol, and 191 for EF-TALFM) are counted as failures in both the internal-energy and joint pass rates. For fragmented outputs, SMILES-based and PoseBusters metrics are computed on the atom-count largest connected component after RDKit sanitization.}
    \label{tbl:pcqm4mv2_generation}
\vspace{0.35em}
\footnotesize
\renewcommand{\arraystretch}{0.98}
\begin{minipage}[t]{0.4\linewidth}
    \vspace{0pt}
    \centering
    \textbf{(a) End-to-end generation results (\%)}
    \vspace{0.25em}

    \setlength{\tabcolsep}{2.5pt}
    \begin{tabular*}{\linewidth}{@{\extracolsep{\fill}}lrrr@{}}
    \toprule
    Metric ($\uparrow$)
    & UAE-3D
    & FlowMol
    & EF-TALFM \\
    \midrule
    \hfill C.
    & 99.60
    & 100.00
    & 92.98 \\
    \hfill S.
    & 83.84
    & 98.72
    & 98.25 \\
    \hfill U.\,\&\,S.
    & 83.81 
    & 98.27
    & 98.21 \\
    \bfseries N.\,\&\,U.\,\&\,S.
    & 76.03
    & 69.83
    & \bfseries 94.33 \\
    \bottomrule
    \end{tabular*}
\end{minipage}
\hfill
\begin{minipage}[t]{0.55\linewidth}
    \vspace{0pt}
    \centering
    \textbf{(b) PoseBusters results among N.\,\&\,U.\,\&\,S. molecules (\%)}
    \vspace{0.25em}

    \setlength{\tabcolsep}{2.5pt}
    \begin{tabular*}{\linewidth}{@{\extracolsep{\fill}}lrrr@{}}
    \toprule
    Metric ($\uparrow$)
    & UAE-3D
    & FlowMol
    & EF-TALFM \\
    \midrule
    Bond lengths
    & 99.96
    & 100.00
    & 99.79 \\
    Bond angles
    & 99.93
    & 100.00
    & 99.32 \\
    Internal steric clash
    & 99.64
    & 99.90
    & 97.22 \\
    Aromatic-ring flatness
    & 100.00
    & 100.00
    & 99.93 \\
    Double-bond flatness
    & 99.91
    & 100.00
    & 98.09 \\
    Internal energy
    & 99.39
    & 99.83
    & 97.69 \\
    \addlinespace[2pt]
    \textbf{All PoseBusters checks} & 98.91 & \bfseries 99.73 & 93.13 \\
    \bottomrule
    \end{tabular*}
\end{minipage}
\end{table}

\subsection{Baseline comparisons at different checkpoints and sampling-step budgets}\label{appendix:unconditional_additional_baseline}
We present \Cref{tbl:pcqm4mv2_generation_additional} for validity and novelty results from additional checkpoints, showing that extended UAE-3D training yields only modest gains, whereas longer FlowMol training improves validity slightly but reduces novelty.
\begin{table}[htbp]
    \centering
    \caption{
    \textbf{EF-TALFM achieves high PCQM4Mv2 generation validity, substantially higher novelty, and competitive geometric realism.}
    We report connectivity of the reconstructed molecule and sanitization, uniqueness, and novelty rates computed after retaining the atom-count largest connected component of each sampled molecule and applying strict RDKit sanitization.
    All rates are computed over 10,000 generated molecules. C. denotes the fraction of connected molecules; S. the fraction passing strict RDKit sanitization; U.\,\&\,S. the fraction of unique sanitized canonical SMILES; and N.\,\&\,U.\,\&\,S. the fraction of generated molecules that are sanitized, unique, and absent from the training set. For fragmented outputs, SMILES-based and PoseBusters metrics are computed on the atom-count largest connected component after RDKit sanitization.} 
    \label{tbl:pcqm4mv2_generation_additional}
    \begin{tabular*}{0.9\linewidth}{@{\extracolsep{\fill}}lrrrrrr@{}}
    \toprule
    Metric ($\uparrow$)
    & UAE-3D$^\ast$
    & UAE-3D$^\S$
    & FlowMol$^\dagger$
    & FlowMol$^\ddagger$
    & FlowMol$^\wr$
    & EF-TALFM \\
    \midrule
    \hfill C.  
    & 99.60
    & 99.71
    & 100.00
    & 100.00
    & 100.00
    & 92.98 \\
    \hfill S.  
    & 83.84
    & 85.89
    & 98.72
    & 99.00
    & 98.57
    & 98.25 \\
    \hfill U.\,\&\,S.  
    & 83.81
    & 85.88
    & 98.27
    & 98.49
    & 98.10
    & 98.21 \\
    \hfill N.\,\&\,U.\,\&\,S. 
    & 76.03
    & 76.62
    & 69.83
    & 65.45
    & 66.50
    & 94.33 \\
    \bottomrule
    \multicolumn{7}{l}{\small $^\ast$best validation checkpoint in 1M steps on latents; 100 DDPM denoising steps;}\\
    \multicolumn{7}{l}{\small $^\S$best validation checkpoint in 2M steps on latents; 500 DDPM denoising steps;}\\
    \multicolumn{7}{l}{\small $^\dagger$best validation checkpoint in 100K steps; 100 Euler integration steps;}\\
    \multicolumn{7}{l}{\small $^\ddagger$best validation checkpoint in 100K steps; 250 Euler integration steps;}\\
    \multicolumn{7}{l}{\small $^\wr$best validation checkpoint in 300K steps; 100 Euler integration steps.}\\
    \bottomrule
    \end{tabular*}
\end{table}

\subsection{Generator-only sampling times}\label{appendix:unconditional_generator_sampling_times}
For sampling, \Cref{fig:generation_compute_efficiency_summary}(b--c) reports the time required to generate $10{,}000$ molecules end-to-end, including generator sampling, reconstruction, and pose relaxation.
\Cref{fig:pcqm4mv2_sampling_time_additional} separately compares generator sampling time for the same number of molecules.
EF-TALFM requires 4 sec using the adaptive fifth-order Dormand--Prince solver implemented in \texttt{torchdiffeq} \citep{torchdiffeq}.
UAE-3D requires 30 sec with 100 DDPM denoising steps and 150 sec with 500 steps, whereas FlowMol requires 141 sec with 100 Euler integration steps.
Thus, for generator sampling alone, EF-TALFM achieves approximate speedups of $7.50\times$ and $37.50\times$ over UAE-3D with 100 and 500 steps, respectively, and $35.25\times$ over FlowMol.

\begin{figure}[htbp]
    \centering
    \includegraphics[width=0.5\linewidth]{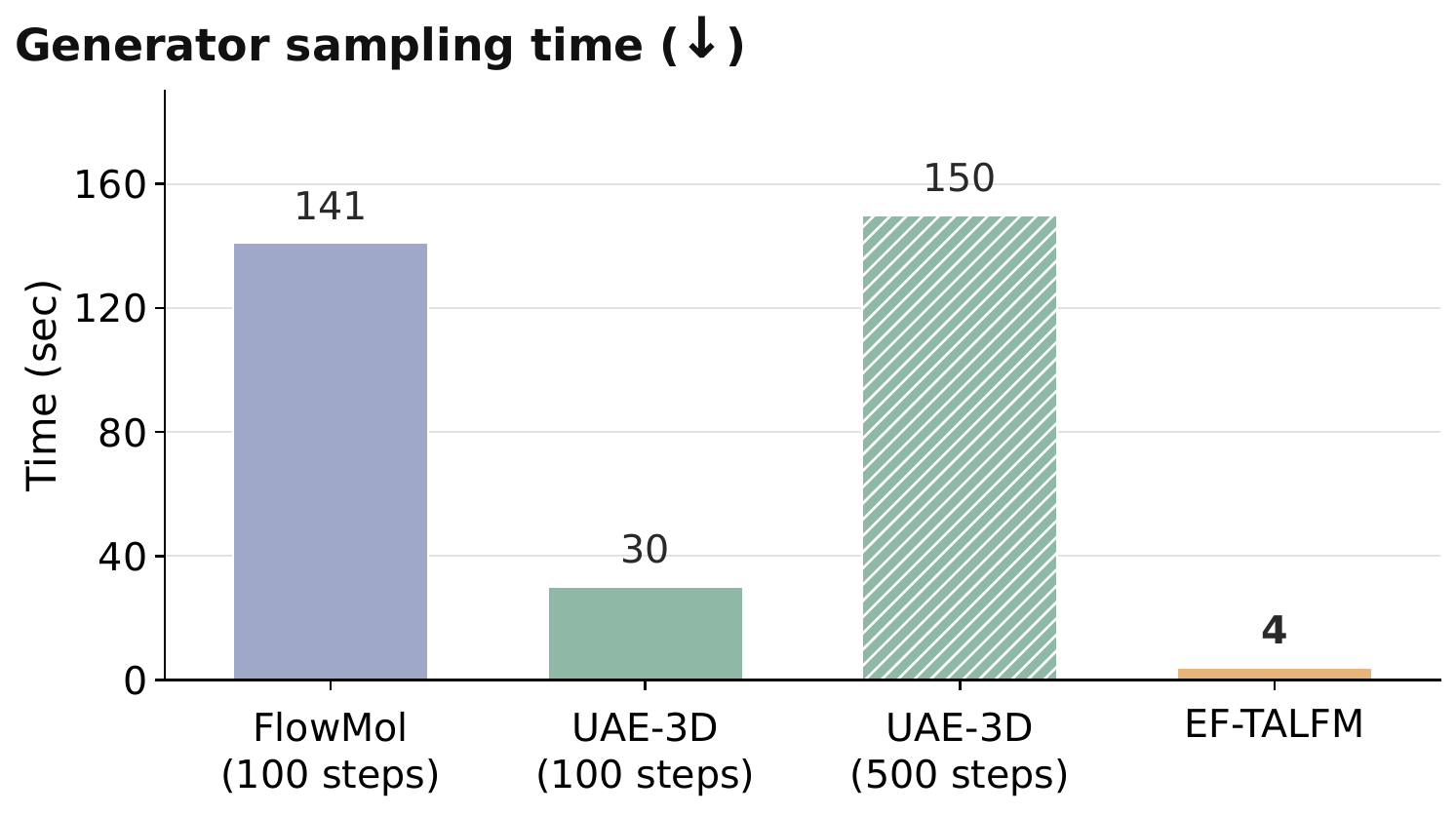}
    \caption{\textbf{EF-TALFM substantially reduces generator sampling time.} Generator-only sampling time for $10{,}000$ molecules. EF-TALFM uses the adaptive fifth-order Dormand--Prince solver, UAE-3D uses a DDPM sampler, and FlowMol uses Euler integration.}
    \label{fig:pcqm4mv2_sampling_time_additional}
\end{figure}

\subsection{Training time and exposure throughput}\label{appendix:unconditional_training_cost}
\Cref{tbl:gradient_evals} provides the optimizer-update budgets, training-exposure proxies and stage-level wall-clock times underlying \Cref{fig:generation_compute_efficiency_summary}d,e.
Training-exposure throughput is total exposure divided by total training time; the stage contributions in \Cref{fig:generation_compute_efficiency_summary}e use the same total time as the denominator.

\begin{table}[htbp]
    \centering 
    \caption{
    \textbf{Optimizer-update budgets, training exposure and training time.}
    EF-TALFM and UAE-3D use two-stage latent training, so training exposure is counted as the number of molecule examples processed at each stage, including repeated appearances across optimizer updates.
    FlowMol instead trains directly in molecular space using sampled edge-level units. Because these units are not directly interchangeable, the reported exposure is a data-volume proxy rather than a measure of total compute.
    }
    \label{tbl:gradient_evals}
    \footnotesize
    \renewcommand{\arraystretch}{1.12}
    \begin{tabular*}{\linewidth}{@{\extracolsep{\fill}}lrrr@{}}
    \toprule
    Method & Optimizer updates & Approx. exposure$^\ast$ & Training time \\
    \midrule
    EF-TALFM (\textit{Total}) & $1.4\,$M & $1,228.8\,$M & $882$ min\\
    \cmidrule{2-4}
    \quad\textit{Variational autoencoder} & $0.4\,$M & $204.8\,$M & $418$ min \\
    \quad\textit{Latent model} & $1.0\,$M & $1,024.0\,$M & $464$ min \\
    \midrule
    UAE-3D (\textit{Total}) & $1.4\,$M & $716.0\,$M & $1,644$ min \\
    \cmidrule{2-4}
    \quad\textit{Variational autoencoder} & $0.4\,$M & $204.8\,$M & $464$ min\\
    \quad\textit{Latent model} & $1.0\,$M & $512.0\,$M & $1,180$ min \\
    \midrule
    FlowMol & $0.1\,$M & $\sim 10.7\,$M & $2,400$ min \\
    \bottomrule
    \end{tabular*}
    \par\vspace{0.5em}
    \begin{minipage}{\linewidth}
    \fontsize{9}{11}\selectfont
    $^\ast$For EF-TALFM and UAE-3D, exposure is optimizer updates times batch size; one unit is one molecule example processed by the VAE or latent generative model. This count does not measure scalar latent elements, padding overhead, or FLOPs. For FlowMol, the native exposure is edge-level: FlowMol was trained for $0.1$M optimizer updates with $320$K sampled edges per update, or approximately $32.0$B edge presentations.
    The reported $\sim 10.7$M value is a rough molecule-equivalent proxy obtained by normalizing $32.0$B edge presentations by FlowMol's sampler estimate of $3$K sampled edges per molecule.
    \end{minipage}
\end{table}
\FloatBarrier

\clearpage

\section[Property-conditioned generation: protocols and supporting results]{Property-conditioned generation: protocols and\\supporting results}\label{appendix:conditional}
\subsection{Experimental setup and evaluation}\label{appendix:conditional_experimental_setup}

\subsubsection{Dataset preprocessing and conditioning targets}\label{appendix:conditional_data_targets}
To provide complete molecular inputs for reference DFT calculations using \texttt{Psi4}, we generate molecules using an all-atom representation.
Although the heavy-atom representation also predicts the number of hydrogens attached to each heavy atom, the explicit hydrogen atoms and their three-dimensional coordinates must subsequently be constructed.
All-atom generation directly models the hydrogen atoms and their positions, reducing reliance on post hoc hydrogen placement and the associated uncertainty in preparing molecular structures for DFT evaluation.
Therefore, we train EF-TALFM on PCQM4Mv2 using an all-atom molecular representation, with property supervision enabled during EF-TAVAE training and the latent flow conditioned on HOMO--LUMO gaps.

Closed-shell species constitute the dominant population in PCQM4Mv2. To maintain a consistent electronic-state domain and avoid mixing this population with sparsely represented radical species, whose frontier-orbital energies depend on spin, we restrict both training and validation to molecular graphs containing no explicit RDKit radical electrons. The resulting fixed partition contains 3{,}000{,}981 training molecules and 115{,}302 validation molecules.

\textbf{Conditioning targets.} For property-conditioned generation on the PCQM4Mv2 dataset, target HOMO--LUMO gaps are selected to span the central region of the training-set distribution. Ten target values are evenly spaced between the 10th and 90th percentiles and rounded to the nearest $0.1\,\mathrm{eV}$, yielding $\mathcal{G}=\{4.1,4.6,5.0,5.4,5.8,6.2,6.6,7.0,7.4,7.8\}$ ($\mathrm{eV}$). For each target $y_{\mathrm{target}}\in\mathcal{G}$, 10,000 molecular samples are generated.

\subsubsection{Screening, selection and evaluation populations}\label{appendix:conditional_population_accounting}
The initial population contains $100{,}000$ generated molecular structures, with $10{,}000$ samples at each of the ten conditioning targets.
Among the $92{,}221$ connected molecules, $90{,}807$ reconstructed graphs pass strict sanitization before geometry refinement.
Of the $90{,}807$ sanitized relaxed molecules, $77{,}976$ ($85.9\%$) contain no explicit RDKit radical electrons, as in the training distribution; $77{,}724$ ($85.6\%$) also pass the PoseBusters geometric validity checks and form the screened population $\mathcal{M}_{\mathrm{gen}}$.
Both percentages use the $90{,}807$ sanitized relaxed molecules as their denominator.

Within each target, we rank screened molecules by their internal-readout error $e_{\mathrm{sur}}$ and retain the best $30\%$.
Rounding the retained count independently within each target and then pooling yields $23{,}315$ selected molecules, denoted $\widehat{\mathcal{M}}_{\mathrm{gen}}$.
For each error threshold, the reported hit rate is the number of hits divided by the corresponding screened or selected population size, $77{,}724$ or $23{,}315$.
At $0.1\,\mathrm{eV}$, the hit subsets $\mathcal{M}_{\mathrm{gen}}^\ast$ and $\widehat{\mathcal{M}}_{\mathrm{gen}}^\ast$ contain $19{,}453$ and $12{,}228$ samples, respectively; at $0.2\,\mathrm{eV}$, the corresponding hit counts are $34{,}919$ and $18{,}764$ (\Cref{tbl:psi4_accuracy_ablation_study}).
Deduplicating the $0.1\,\mathrm{eV}$ hits separately within each target gives $19{,}338$ unique screened hits and $12{,}154$ unique selected hits after pooling, as used in the novelty and diversity analyses below.

\subsubsection{Novelty and diversity metrics} \label{appendix:conditional_novelty_and_diversity}
Novelty and diversity are assessed within targets before pooling.
\textbf{Uniqueness and exact-match training-set novelty.}
For each conditioning target \(g\in\mathcal{G}\), let
\(\mathcal{M}_{g}\) denote the generated molecules assigned to that target,
\(\mathcal{U}_{g}\) their set of unique molecules under the exact-match key defined below, and
\(\mathcal{T}_{g}\) the set of unique training molecules satisfying the
same property tolerance,
\[
\mathcal{T}_{g}
=
\left\{
\mathcal{M}\in\mathcal{D}_{\mathrm{train}}:
\left|y_{\mathrm{train}}(\mathcal{M})-g\right|<0.1~\mathrm{eV}
\right\}.
\]
Exact molecular identity is represented by canonical SMILES
after removing explicit hydrogen atoms while retaining formal charges and bond
orders. Each resulting string is used as an exact-match key. The pooled sample
count, uniqueness rate, and exact-match training-set novelty among unique
molecules are
\[
N=\sum_{g\in\mathcal{G}}|\mathcal{M}_{g}|,
\qquad
\mathrm{U.}
=
\frac{\sum_g|\mathcal{U}_{g}|}
     {\sum_g|\mathcal{M}_{g}|},
\qquad
R_{\mathrm{exact}}
=
\frac{\sum_g|\mathcal{U}_{g}\setminus\mathcal{T}_{g}|}
     {\sum_g|\mathcal{U}_{g}|}.
\]
Here, set membership is evaluated by exact equality of the canonical-SMILES
keys. Unique molecules are formed separately within each target before pooling;
the same molecule generated under two targets therefore contributes once
to each target-specific set.

\textbf{Tanimoto similarity.}
Let \(\mathbf{f}(m)\) denote the 2,048-bit ECFP4 fingerprint of molecule \(m\),
calculated with radius 2, bond types included, and chirality excluded. We use
ECFP4 to measure analogue-level graph similarity independently of 3D pose.
Its radius-2 circular neighborhoods encode local environments extending up to
two bonds from each atom; the ``4'' denotes the corresponding diameter. The
Tanimoto similarity between molecules \(\mathcal{M}_i\) and \(\mathcal{M}_j\) is
\[
\tau(\mathcal{M}_i,\mathcal{M}_j)
=
\frac{
\mathbf{f}(\mathcal{M}_i)^\mathsf{T}\mathbf{f}(\mathcal{M}_j)
}{
\|\mathbf{f}(\mathcal{M}_i)\|_1
+
\|\mathbf{f}(\mathcal{M}_j)\|_1
-
\mathbf{f}(\mathcal{M}_i)^\mathsf{T}\mathbf{f}(\mathcal{M}_j)
}.
\]
For each unique generated molecule \(\mathcal{M}\in\mathcal{U}_{g}\), its
exact maximum similarity to the target-matched training reference is
\[
s_{\max}(\mathcal{M})
=
\max_{\mathcal{M}^\prime\in\mathcal{T}_{g}}
\tau(\mathcal{M},\mathcal{M}^\prime).
\]
The pooled maximum-similarity vectors are concatenated across targets. The table reports their median, interquartile range
\[
\mathrm{IQR}(s_{\max})
=
Q_{0.75}(s_{\max})-Q_{0.25}(s_{\max}),
\]
and the fraction below a threshold \(\delta\in\{0.4,0.8\}\),
\[
R_{\delta}
=
\frac{
\sum_g\sum_{\mathcal{M}\in\mathcal{U}_{g}}
\mathbb{I}\!\left[s_{\max}(\mathcal{M})<\delta\right]
}{
\sum_g|\mathcal{U}_{g}|
}.
\]

\textbf{Scaffold novelty.}
Let \(\mathcal{S}(\mathcal{A})\) denote the set of distinct nonempty
Bemis--Murcko scaffolds occurring in molecular set \(\mathcal{A}\). Scaffold
novelty is calculated target-wise and then pooled as
\[
R_{\mathrm{scaffold}}
=
\frac{
\sum_g
\left|
\mathcal{S}(\mathcal{U}_{g})
\setminus
\mathcal{S}(\mathcal{T}_{g})
\right|
}{
\sum_g
\left|
\mathcal{S}(\mathcal{U}_{g})
\right|
}.
\]
Acyclic molecules have empty Bemis--Murcko scaffolds and therefore do not
contribute to this calculation. Because scaffolds are collected separately
within each target, the same scaffold can contribute once for each target in
which it occurs. The pooled rate therefore describes novel target--scaffold
pairs and does not globally deduplicate scaffolds across the ten targets.

\textbf{Internal-diversity.}
For internal-diversity analysis, the pooled number of unique molecules and
the number of within-target unordered molecular pairs are
\[
N_{\mathrm{unique}}
=
\sum_g|\mathcal{U}_{g}|,
\qquad
P
=
\sum_g\binom{|\mathcal{U}_{g}|}{2}.
\]
The pooled mean pairwise similarity is
\[
\overline{\tau}_{\mathrm{pair}}
=
\frac{
\displaystyle
\sum_g
\sum_{\{\mathcal{M}_i,\mathcal{M}_j\}\subset\mathcal{U}_{g}}
\tau(\mathcal{M}_i,\mathcal{M}_j)
}{
\displaystyle
\sum_g\binom{|\mathcal{U}_{g}|}{2}
},
\]
and internal diversity is defined as
\[
D_{\mathrm{int}}
=
1-\overline{\tau}_{\mathrm{pair}}.
\]
Thus, only pairs belonging to the same target are compared; no cross-target
molecular pairs are introduced.

For each target \(g\) and resampling replicate \(r\), the training control is
sampled without replacement and size-matched to the unique generated
molecules for that target:
\[
\mathcal{C}_{g}^{(r)}
\subseteq
\mathcal{T}_{g},
\qquad
\left|\mathcal{C}_{g}^{(r)}\right|
=
\left|\mathcal{U}_{g}\right|.
\]
Applying this construction to \(\mathcal{M}_{\mathrm{gen}}^\ast\) and
\(\widehat{\mathcal{M}}_{\mathrm{gen}}^\ast\) gives
\(\widetilde{\mathcal{M}}_{\mathrm{train}}^\ast\) and
\(\widehat{\mathcal{M}}_{\mathrm{train}}^\ast\), respectively.
For each replicate, the pooled training similarity
\(\overline{\tau}_{\mathrm{train}}^{(r)}\) uses the same pair-count-weighted
expression as the generated cohort: target-wise mean similarities are weighted
by their numbers of within-target unordered pairs, and no cross-target pairs
are introduced. Over \(R=20\) target-stratified resamples, the reported
training-control value is
\[
\mu_{\mathrm{train}}
=
\frac{1}{R}
\sum_{r=1}^{R}
\overline{\tau}_{\mathrm{train}}^{(r)},
\qquad
\sigma_{\mathrm{train}}
=
\sqrt{
\frac{1}{R-1}
\sum_{r=1}^{R}
\left(
\overline{\tau}_{\mathrm{train}}^{(r)}
-\mu_{\mathrm{train}}
\right)^2
},
\]
reported as
\(\mu_{\mathrm{train}}\pm\sigma_{\mathrm{train}}\). The corresponding
training internal diversity is
\(1-\mu_{\mathrm{train}}\), with the same sample standard deviation.

\subsubsection{Training-density analysis in UMAP space}\label{appendix:conditional_density_analysis}
We re-encode generated molecules with EF-TAVAE and apply a shared uniform manifold approximation and projection (UMAP)~\citep{mcinnes2018umap} mapping of the penultimate molecule-path property representation $\mathbf r$, fitted to a fixed sample of 100{,}000 training molecules.
We estimate local training density from the full training population in the shared UMAP projection using a Gaussian-smoothed two-dimensional histogram.
Each generated candidate is assigned the percentile of its interpolated local density relative to the densities at the training coordinates, so low percentiles indicate sparsely populated regions of this UMAP projection.
Using the same density reference for both populations, we divide candidates into ten percentile bins and compute the fraction satisfying each DFT-error threshold within each bin, pooling candidates across targets.

\clearpage

\subsection{PCA analysis of the learned property representation} \label{appendix:pca_visualization}

\Cref{fig:pca_appendix} shows the same screened and selected populations in the first two principal components of $\mathbf r$, providing a linear view of the leading variance directions that complements the nonlinear, neighborhood-based UMAP view in \Cref{fig:internal_selection_and_umap}(d).
Consistent with the UMAP analysis, the $0.1$ eV hits occupy multiple regions of the projection, and selection increases the hit rate in every PCA training-density decile (\Cref{fig:pca_appendix}(b)).

\begin{figure}[htbp]
    \centering
    \includegraphics[width=1.0\linewidth]{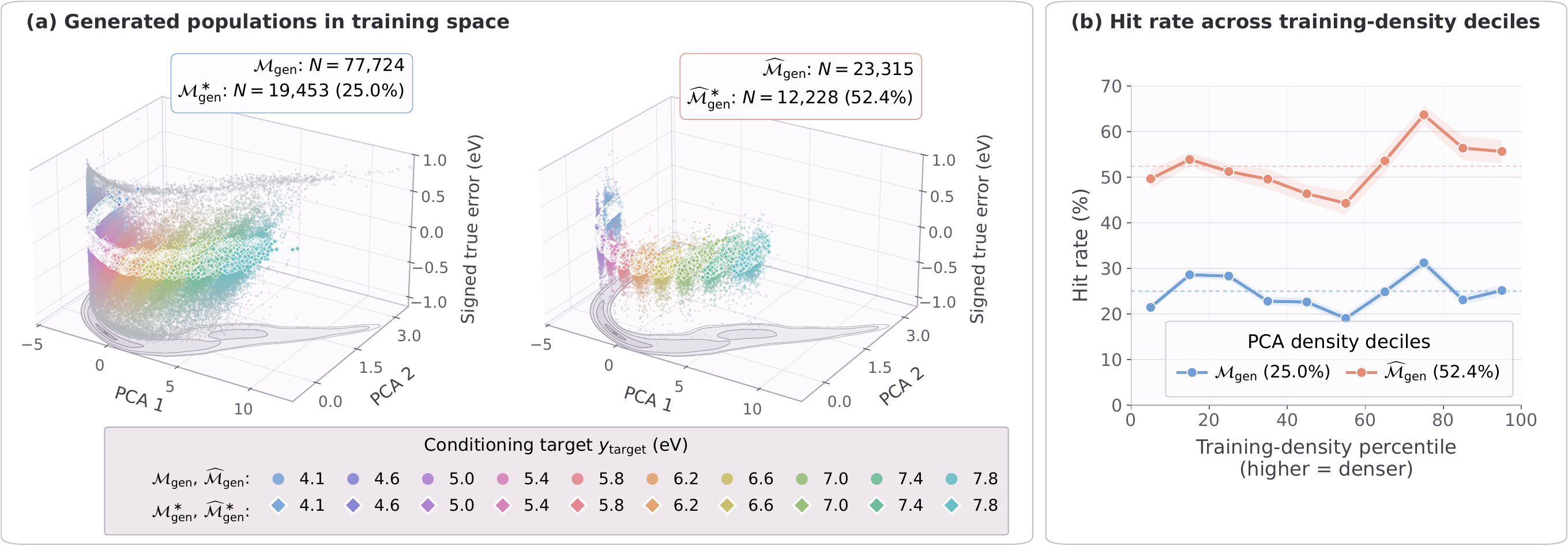}
    \caption{
    \textbf{Internal-readout selection increases DFT-verified hit rates across every training-density decile in the learned property representation.}
    (a) PCA mapping of the penultimate molecule-path property representation $\mathbf r$. It compares the screened relaxed population $\mathcal{M}_{\mathrm{gen}}$ (left) with its selected subset $\widehat{\mathcal{M}}_{\mathrm{gen}}$ (right), obtained by retaining the target-wise best $30\%$ ranked by $e_{\mathrm{sur}}$.
    The signed true-property error, $y_{\mathrm{true}}-y_{\mathrm{target}}$ is clipped at $\pm1.0$ eV. 
    Colors identify the ten conditioning targets.
    Generation hits with $\varepsilon_{\mathrm{true}}<0.1$ are shown as white-outlined diamond-shaped markers, whereas misses are represented by round-shaped markers whose target colors progressively fade toward gray with increasing absolute error.
    The floor shading and contours represent the smoothed density of the full training population; contour boundaries enclose the highest-density regions containing $q\in\{1,10,50,90,99,99.5\}\%$ of the training molecules. 
    The corresponding hit populations are $\mathcal{M}_{\mathrm{gen}}^\ast$ and $\widehat{\mathcal{M}}_{\mathrm{gen}}^\ast$. 
    Panel (b) reports the corresponding hit rates across training-density percentile deciles; shaded bands denote \(95\%\) Wilson intervals and dashed lines indicate overall hit rates.
    }
    \label{fig:pca_appendix}
\end{figure}

\FloatBarrier
\clearpage

\subsection{Selection at different retention thresholds}\label{appendix:selection_at_different_retention_thresholds}
The retention sweep in \Cref{tbl:psi4_accuracy_ablation_study} shows how increasingly stringent readout-based selection trades candidate coverage for higher DFT-verified hit rates; the main comparison uses the 30\% retention setting included in this sweep.

\begin{table}[htbp]
    \centering
    \caption{ 
    \textbf{Internal-readout--based selection at different retention thresholds.} 
    During selection, the screened relaxed molecules within each target value are ranked by the internal-readout error $e_{\mathrm{sur}}$, and the indicated ``Best'' fraction is retained before pooling across targets.
    Means are reported as mean (std) of the molecule-wise absolute true-property errors; both quantities are in eV.
    For a threshold $\delta$, Count is the number of retained molecules satisfying
    $\varepsilon_{\mathrm{true}}<\delta$, and Rate is Count as a percentage of the
    retained-set size listed in the $N$ column, not of the initial 100{,}000.
    Retained counts are rounded independently within each target before pooling.
    }
    \label{tbl:psi4_accuracy_ablation_study}
     
    \footnotesize
    \setlength{\tabcolsep}{3pt}
    \renewcommand{\arraystretch}{1.12}
    
    \begin{tabular*}{\linewidth}{
        @{\extracolsep{\fill}}l
        S[
            table-format=5.0,
            group-separator={,},
            group-minimum-digits=4
        ]
        c
        S[table-format=1.3]
        S[table-format=2.1]
        S[
            table-format=5.0,
            group-separator={,},
            group-minimum-digits=4
        ]
        S[table-format=2.1]
        S[
            table-format=5.0,
            group-separator={,},
            group-minimum-digits=4
        ]
        @{}
    }
    \toprule
    &
    &
    \multicolumn{2}{c}{True error \(\varepsilon_{\mathrm{true}}\) (eV)}
    &
    \multicolumn{2}{c}{\(\varepsilon_{\mathrm{true}}<0.1\) eV}
    &
    \multicolumn{2}{c}{\(\varepsilon_{\mathrm{true}}<0.2\) eV}
    \\
    \cmidrule(lr){3-4}
    \cmidrule(lr){5-6}
    \cmidrule(lr){7-8}
    
    Retained set
    & \multicolumn{1}{c}{\(N\)}
    & \multicolumn{1}{c}{Mean (Std) \(\downarrow\)}
    & \multicolumn{1}{c}{Median \(\downarrow\)}
    & \multicolumn{1}{c}{Rate (\%) \(\uparrow\)}
    & \multicolumn{1}{c}{Count \(\uparrow\)}
    & \multicolumn{1}{c}{Rate (\%) \(\uparrow\)}
    & \multicolumn{1}{c}{Count \(\uparrow\)}
    \\
    \midrule
    
    \multicolumn{8}{@{}l}{\textit{Screened population}}\\[2pt]
    
    \hfill $\mathcal{M}_{\mathrm{gen}}$
    & 77724
    & 0.350 (0.385)
    & 0.232
    & 25.0
    & 19453
    & 44.9
    & 34919
    \\
    
    \midrule
    
    \multicolumn{8}{@{}l}{%
    \textit{Selection with \(e_{\mathrm{sur}}\) on screened refined structures}}\\[2pt]
    
    \hfill Best 10\%
    & 7774
    & 0.119 (0.140)
    & 0.084
    & 56.9
    & 4424
    & 83.1
    & 6457
    \\
    
    \hfill Best 20\%
    & 15545
    & 0.122 (0.144)
    & 0.087
    & 55.5
    & 8620
    & 82.2
    & 12780
    \\
    
    \hfill Best 30\%
    & 23315
    & 0.130 (0.157)
    & 0.094
    & 52.4
    & 12228
    & 80.5
    & 18764
    \\
    
    \hfill Best 50\%
    & 38863
    & 0.152 (0.161)
    & 0.115
    & 44.2
    & 17196
    & 73.9
    & 28707
    \\
    
    \bottomrule
    \end{tabular*}
\end{table}

\FloatBarrier
\clearpage

\subsection{Per-target novelty and diversity of DFT-verified hits}\label{appendix:condition_ablation_study}

\begin{table}[htbp]
    \centering
    \caption{
    \textbf{Per-target and pooled novelty of DFT-verified hits.}
    The screened hits, $\mathcal{M}_{\mathrm{gen}}^\ast$, and selected hits,
    $\widehat{\mathcal{M}}_{\mathrm{gen}}^\ast$, both satisfy
    $|y_{\mathrm{true}}-y_{\mathrm{target}}|<0.1\,\mathrm{eV}$.
    Selection retains the best 30\% of screened candidates within each target ranked by $e_{\mathrm{sur}}$.
    We report uniqueness, exact-match novelty, maximum ECFP4 Tanimoto similarity to the target-matched training subset, and scaffold novelty.
    All metrics and pooled rows follow the definitions in \Cref{appendix:conditional_novelty_and_diversity}.
    }
    \label{tab:generated_novelty_internal_diversity_ten_target}

    \footnotesize
    \setlength{\tabcolsep}{4pt}
    \renewcommand{\arraystretch}{1.12}

    \sisetup{
        detect-weight=true,
        group-separator={,},
        group-digits=integer,
        group-minimum-digits=4
    }

    \begin{tabular*}{\linewidth}{
        @{\extracolsep{\fill}}l
        S[table-format=5.0, group-separator={,}, group-minimum-digits=4]
        S[table-format=2.1]
        S[table-format=2.1]
        c
        S[table-format=2.1]
        S[table-format=2.1]
        S[table-format=2.1]
        @{}
    }
    \toprule
    &
    \multicolumn{3}{c}{}
    &
    \multicolumn{3}{c}{\shortstack{Exact maximum training-set\\ECFP4 Tanimoto}}
    &
    \multicolumn{1}{c}{}
    \\
    \cmidrule(lr){2-4}
    \cmidrule(lr){5-7}

    \multicolumn{1}{c}{$y_{\mathrm{target}}$}
    & \multicolumn{1}{c}{Sample $N$}
    & \multicolumn{1}{c}{U.}
    & \multicolumn{1}{c}{U.\&\,N.}
    & \multicolumn{1}{c}{Median (IQR)}
    & \multicolumn{1}{c}{$<0.4$}
    & \multicolumn{1}{c}{$<0.8$}
    & \multicolumn{1}{c}{\shortstack{Scaffold\\novelty}}
    \\
    \multicolumn{1}{c}{(eV)}
    &
    & \multicolumn{1}{c}{(\%)}
    & \multicolumn{1}{c}{(\%)}
    &
    & \multicolumn{1}{c}{(\%)}
    & \multicolumn{1}{c}{(\%)}
    & \multicolumn{1}{c}{(\%)}
    \\
    \midrule
    \multicolumn{8}{@{}l}{
        \textit{Screened hits, $\mathcal{M}_{\mathrm{gen}}^\ast$}
    }
    \\
    \addlinespace[2pt]

    4.1 & 1070 & 99.9 & 99.4 & 0.452 (0.139) & 26.9 & 99.2 & 53.1 \\
    4.6 & 1230 & 99.7 & 98.4 & 0.500 (0.156) & 16.1 & 97.7 & 36.4 \\
    5.0 & 1746 & 99.9 & 98.3 & 0.538 (0.151) &  8.9 & 96.8 & 31.1 \\
    5.4 & 2048 & 99.6 & 97.1 & 0.574 (0.158) &  5.0 & 94.9 & 31.1 \\
    5.8 & 2696 & 99.5 & 96.3 & 0.611 (0.166) &  3.4 & 93.2 & 29.4 \\
    6.2 & 2140 & 99.3 & 97.3 & 0.565 (0.159) &  5.9 & 94.7 & 34.1 \\
    6.6 & 2064 & 99.1 & 96.3 & 0.578 (0.196) &  9.1 & 90.2 & 39.2 \\
    7.0 & 2261 & 99.6 & 98.8 & 0.548 (0.167) &  9.8 & 95.3 & 39.1 \\
    7.4 & 2220 & 98.8 & 97.9 & 0.543 (0.167) &  8.2 & 94.0 & 36.7 \\
    7.8 & 1978 & 99.1 & 98.6 & 0.552 (0.173) &  8.5 & 94.3 & 50.6 \\
    \midrule
    \textbf{Pooled}
    & \bfseries 19453
    & \bfseries 99.4
    & \bfseries 97.7
    & \bfseries 0.556 (0.173)
    & \bfseries 8.9
    & \bfseries 94.6
    & \bfseries 37.6
    \\
    \midrule
    \addlinespace[2pt]
    \multicolumn{8}{@{}l}{
        \textit{Selected hits, $\widehat{\mathcal{M}}_{\mathrm{gen}}^\ast$}
    }
    \\
    \addlinespace[2pt]

    4.1 &  743 & 99.9 & 99.3 & 0.469 (0.141) & 22.5 & 98.9 & 48.3 \\
    4.6 &  794 & 99.6 & 98.0 & 0.514 (0.153) & 12.3 & 97.2 & 28.5 \\
    5.0 & 1147 & 99.9 & 98.4 & 0.545 (0.148) &  7.2 & 96.8 & 27.2 \\
    5.4 & 1308 & 99.5 & 96.5 & 0.581 (0.156) &  4.1 & 94.8 & 29.2 \\
    5.8 & 1705 & 99.4 & 96.0 & 0.628 (0.152) &  1.9 & 92.5 & 23.4 \\
    6.2 & 1364 & 99.3 & 97.0 & 0.577 (0.149) &  3.7 & 94.4 & 29.0 \\
    6.6 & 1311 & 98.9 & 95.0 & 0.605 (0.210) &  5.9 & 87.2 & 36.1 \\
    7.0 & 1399 & 99.5 & 99.1 & 0.559 (0.151) &  7.1 & 95.7 & 37.8 \\
    7.4 & 1266 & 98.7 & 97.5 & 0.554 (0.171) &  7.4 & 93.3 & 33.8 \\
    7.8 & 1191 & 99.5 & 98.2 & 0.559 (0.164) &  7.0 & 94.2 & 44.6 \\
    \midrule
    \textbf{Pooled}
    & \bfseries 12228
    & \bfseries 99.4
    & \bfseries 97.4
    & \bfseries 0.568 (0.171)
    & \bfseries 6.8
    & \bfseries 94.1
    & \bfseries 33.5
    \\
    \bottomrule
    \end{tabular*}
    \end{table}

    \begin{table}[p]
    \centering
    \caption{
    \textbf{Per-target and pooled internal diversity of DFT-verified hits.}
    Both hit sets satisfy $|y_{\mathrm{true}}-y_{\mathrm{target}}|<0.1\,\mathrm{eV}$; selection retains the best 30\% of screened candidates within each target ranked by $e_{\mathrm{sur}}$.
    Mean similarity averages ECFP4 Tanimoto similarity over within-target unordered pairs; internal diversity is one minus this mean.
    The training control is target- and size-matched to unique screened hits; its values are means $\pm$ sample standard deviations over 20 resamples.
    Pooled means use within-target pair-count weights (\Cref{appendix:conditional_novelty_and_diversity}).
    }
    \label{tab:generated_internal_diversity_ten_target}
    \footnotesize
    \setlength{\tabcolsep}{4pt}
    \renewcommand{\arraystretch}{1.12}
    \sisetup{detect-weight=true,group-separator={,},group-digits=integer,group-minimum-digits=4}
    \begin{tabular*}{\linewidth}{@{\extracolsep{\fill}}lrr
        S[table-format=1.4(2),separate-uncertainty=true]
        S[table-format=1.4(2),separate-uncertainty=true]@{}}
    \toprule
    \multicolumn{1}{c}{$y_{\mathrm{target}}$ (eV)}
    & \multicolumn{1}{c}{Unique $N$}
    & \multicolumn{1}{c}{Within-target pairs}
    & \multicolumn{1}{c}{Mean similarity}
    & \multicolumn{1}{c}{Internal diversity} \\
    \midrule
    \multicolumn{5}{@{}l}{\textit{Screened hits, $\mathcal{M}_{\mathrm{gen}}^\ast$}}\\
    \addlinespace[2pt]
    4.1 & \num{1069} & \num{570846} & 0.1065 & 0.8935 \\
    4.6 & \num{1226} & \num{750925} & 0.1043 & 0.8957 \\
    5.0 & \num{1745} & \num{1521640} & 0.1058 & 0.8942 \\
    5.4 & \num{2040} & \num{2079780} & 0.1136 & 0.8864 \\
    5.8 & \num{2682} & \num{3595221} & 0.1350 & 0.8650 \\
    6.2 & \num{2126} & \num{2258875} & 0.1130 & 0.8870 \\
    6.6 & \num{2046} & \num{2092035} & 0.1054 & 0.8946 \\
    7.0 & \num{2251} & \num{2532375} & 0.1098 & 0.8902 \\
    7.4 & \num{2193} & \num{2403528} & 0.1137 & 0.8863 \\
    7.8 & \num{1960} & \num{1919820} & 0.1255 & 0.8745 \\
    \midrule
    \textbf{Pooled} & \bfseries \num{19338} & \bfseries \num{19725045} & \bfseries 0.1161 & \bfseries 0.8839 \\
    \midrule
    \addlinespace[3pt]
    \multicolumn{5}{@{}l}{\textit{Selected hits, $\widehat{\mathcal{M}}_{\mathrm{gen}}^\ast$}}\\
    \addlinespace[2pt]
    4.1 & \num{742} & \num{274911} & 0.1085 & 0.8915 \\
    4.6 & \num{791} & \num{312445} & 0.1059 & 0.8941 \\
    5.0 & \num{1146} & \num{656085} & 0.1081 & 0.8919 \\
    5.4 & \num{1302} & \num{846951} & 0.1163 & 0.8837 \\
    5.8 & \num{1695} & \num{1435665} & 0.1457 & 0.8543 \\
    6.2 & \num{1355} & \num{917335} & 0.1236 & 0.8764 \\
    6.6 & \num{1296} & \num{839160} & 0.1158 & 0.8842 \\
    7.0 & \num{1392} & \num{968136} & 0.1161 & 0.8839 \\
    7.4 & \num{1250} & \num{780625} & 0.1190 & 0.8810 \\
    7.8 & \num{1185} & \num{701520} & 0.1278 & 0.8722 \\
    \midrule
    \textbf{Pooled} & \bfseries \num{12154} & \bfseries \num{7732833} & \bfseries 0.1224 & \bfseries 0.8776 \\
    \midrule
    \addlinespace[3pt]
    \multicolumn{5}{@{}l}{\textit{Training control, $\widetilde{\mathcal{M}}_{\mathrm{train}}^\ast$}}\\
    \addlinespace[2pt]
    4.1 & \num{1069} & \num{570846} & 0.0835(14) & 0.9165(14) \\
    4.6 & \num{1226} & \num{750925} & 0.0839(10) & 0.9161(10) \\
    5.0 & \num{1745} & \num{1521640} & 0.0848(8) & 0.9152(8) \\
    5.4 & \num{2040} & \num{2079780} & 0.0854(9) & 0.9146(9) \\
    5.8 & \num{2682} & \num{3595221} & 0.0857(6) & 0.9143(6) \\
    6.2 & \num{2126} & \num{2258875} & 0.0848(6) & 0.9152(6) \\
    6.6 & \num{2046} & \num{2092035} & 0.0847(7) & 0.9153(7) \\
    7.0 & \num{2251} & \num{2532375} & 0.0827(7) & 0.9173(7) \\
    7.4 & \num{2193} & \num{2403528} & 0.0820(8) & 0.9180(8) \\
    7.8 & \num{1960} & \num{1919820} & 0.0841(7) & 0.9159(7) \\
    \midrule
    \textbf{Pooled} & \bfseries \num{19338} & \bfseries \num{19725045} & \bfseries 0.0843(2) & \bfseries 0.9157(2) \\
    \bottomrule
    \end{tabular*}
\end{table}

\FloatBarrier
\clearpage

\section{Methods}\label{appendix:methods} 
\subsection{Training and sampling algorithms of EF-TALFM}\label{appendix:training_sampling_algorithms}
\begin{algorithm}[H]
    \caption{Training algorithm of EF-TALFM}\label{algo:training}
    \begin{algorithmic}[1]
        \INPUT Molecule data in the form of $ \mathcal{M} =\left( \mathbf{u}, \{(\mathbf{x}_i,\mathbf{h}_i)\}_{i=1}^{n}  \right)$,  
        initialized encoder $\bm f_{\bm \phi}$, decoder $\bm f_{\bm \psi}$, $\mathbf z^{(0)}$, \texttt{MolEnc} with parameters $\bm\xi_m$, \texttt{AtomEnc} with $\bm\xi_a$, \texttt{LatEnc} with $\bm\xi_z$, \texttt{Atom2Enc} with $\tilde{\bm\xi}_a$, \texttt{MolDec} with $\bm\zeta_m$, \texttt{AtomDec} with $\bm\zeta_a$, flow matching neural networks $\bm f_{\bm\theta}$; property $\mathbf y$, initialized $\bm \nu_{\texttt{[PROP]}}^{(0)}$, \texttt{PropDec} with $\bm\zeta_y$, if auxiliary property-prediction head enabled.
        \STATE \textbf{First Stage: Variational Autoencoder Training}
        \REPEAT
        \STATE Sample $\mathcal{M} =\left( \mathbf{u}, \{(\mathbf{x}_i,\mathbf{h}_i)\}_{i=1}^{n}  \right)$
        \STATE $\{( \mathbf x_i, 
         \mathbf h_i)\}_{i=1}^n \gets \texttt{CanonicalAlignment}(\{(\mathbf{x}_i,\mathbf{h}_i)\}_{i=1}^{n})$
        \STATE Append the nonphysical end-of-molecule token $(\mathbf x_{n+1}, 
        \mathbf h_{n+1})$ to $\{( \mathbf x_i, 
         \mathbf h_i)\}_{i=1}^n$
        \STATE Construct $\mathbf e_i^{(0)}\gets \texttt{MolEnc}_{\bm \xi_m}(\mathbf u) + \texttt{AtomEnc}_{\bm \xi_a}(\mathbf x_i, \mathbf h_i)$, $i=1,\ldots,n+1$
        \STATE $[(\mathbf z_\mu, \mathbf z_\sigma), \mathbf e_1^{(L)},\cdots, \mathbf e_{n+1}^{(L)}]   \gets \bm f_{\bm\phi}([\mathbf z^{(0)}, \mathbf e_1^{(0)},\cdots, \mathbf e_{n+1}^{(0)}])$  
        \STATE Sample $\mathbf z \gets \mathbf z_\mu + \mathbf z_{\sigma} \odot \bm\varepsilon $ with $\bm\varepsilon\sim\mathcal{N}(\mathbf 0, \mathbf I)$
        \STATE Decode $\hat{\mathbf{u}}\gets \texttt{MolDec}_{\bm\zeta_m}(\mathbf z)$
        \STATE Encode $\bm \nu_0^{(0)} \gets \texttt{LatEnc}_{\bm\xi_z}(\mathbf z)$ 
        \FOR{$i$ in $1, 2, \cdots, n+1$}
        \STATE Construct causal attention matrix $\mathbf A_{\mathrm{causal},i}$
        \STATE $\bm \nu_{i-1}^{(L)} \gets f_{\bm \psi}( [\bm \nu_{0}^{(0)},\cdots ,\bm \nu_{i-1}^{(0)}]; \mathbf A_{\mathrm{causal},i})$
        \STATE $(\hat{\mathbf x}_i, \hat{\mathbf h}_i)\gets \texttt{AtomDec}_{\bm\zeta_a}(\bm \nu_{i-1}^{(L)})$
        \STATE $\bm \nu_i^{(0)}\gets \texttt{Atom2Enc}_{\tilde{\bm\xi}_a}(\mathbf x_i, \mathbf h_i)$ \hfill $\triangleright$ \textit{Teacher forcing with ground-truth inputs}
        \ENDFOR
        \IF{auxiliary property-prediction head enabled}
        \STATE Construct molecule-conditioned and latent-only attention matrices, $\mathbf A_{\mathrm{mol}}$ and $\mathbf A_{\mathrm{lat}}$
        \STATE $\bm\nu_{\texttt{[PROP]},\mathrm{lat}}^{(L)}\gets f_{\bm \psi}( [\bm \nu_{0}^{(0)},\bm \nu_{\texttt{[PROP]}}^{(0)},\bm \nu_{1}^{(0)},\cdots ,\bm \nu_{n+1}^{(0)}]; \mathbf A_{\mathrm{lat}})$
        \STATE $\hat{\mathbf y}_{\mathrm{lat}}\gets \texttt{PropDec}_{\bm \zeta_y}(\bm\nu_{\texttt{[PROP]},\mathrm{lat}}^{(L)})$
        \STATE $\bm\nu_{\texttt{[PROP]},\mathrm{mol}}^{(L)}\gets f_{\bm \psi}( [\bm \nu_{0}^{(0)},\bm \nu_{\texttt{[PROP]}}^{(0)},\bm \nu_{1}^{(0)},\cdots ,\bm \nu_{n+1}^{(0)}]; \mathbf A_{\mathrm{mol}})$
        \STATE $\hat{\mathbf y}_{\mathrm{mol}}\gets \texttt{PropDec}_{\bm \zeta_y}(\bm\nu_{\texttt{[PROP]},\mathrm{mol}}^{(L)})$
        \ENDIF
        \STATE Compute the full VAE loss $\mathcal{L}_{\mathrm{VAE}}$ as in \Cref{eq:training_objective}
        \STATE Update $\bm\phi$, $\bm\psi$, $\mathbf z^{(0)}$, $\bm\xi_m$, $\bm\xi_a$, $\bm\xi_z$, $\tilde{\bm\xi}_a$, $\bm\zeta_m$ and $\bm\zeta_a$
        \UNTIL Converged
        \STATE \textbf{Second Stage: Latent Flow Matching Training}
        \STATE Fix parameters $\bm\phi$, $\bm\psi$, $\mathbf z^{(0)}$, $\bm\xi_m$, $\bm\xi_a$, $\bm\xi_z$, $\tilde{\bm\xi}_a$, $\bm\zeta_m$ and $\bm\zeta_a$
        \REPEAT
        \STATE Sample $\mathcal{M} =\left( \mathbf{u}, \{(\mathbf{x}_i,\mathbf{h}_i)\}_{i=1}^{n}  \right)$
        \STATE $\{( \mathbf x_i, 
         \mathbf h_i)\}_{i=1}^n \gets \texttt{CanonicalAlignment}(\{(\mathbf{x}_i,\mathbf{h}_i)\}_{i=1}^{n})$
        \STATE Append the nonphysical end-of-molecule token $(\mathbf x_{n+1}, 
        \mathbf h_{n+1})$ to $\{( \mathbf x_i, 
         \mathbf h_i)\}_{i=1}^n$
        \STATE Construct $\mathbf e_i^{(0)}\gets \texttt{MolEnc}_{\bm \xi_m}(\mathbf u) + \texttt{AtomEnc}_{\bm \xi_a}(\mathbf x_i, \mathbf h_i)$, $i=1,\ldots,n+1$
        \STATE $[(\mathbf z_\mu, \mathbf z_\sigma), \mathbf e_1^{(L)},\cdots, \mathbf e_{n+1}^{(L)}]   \gets \bm f_{\bm\phi}([\mathbf z^{(0)}, \mathbf e_1^{(0)},\cdots, \mathbf e_{n+1}^{(0)}])$
        \STATE Sample $\mathbf z_{\varepsilon}\sim\mathcal{N}(\mathbf 0,\mathbf I)$ and $t\sim\mathrm{Unif}(0,1)$
        \STATE Interpolate $\mathbf z_t \gets (1-t)\, \mathbf z_{\varepsilon} + t\,\mathbf z_{\mu}$
        \STATE Target vector field $\bm v \gets \mathbf z_{\mu}-\mathbf z_{\varepsilon}$
        \STATE Estimated vector field $\hat{\bm v}\gets \bm f_{\bm \theta}(\mathbf z_t, t, \mathbf y)$, omitting $\mathbf y$ if unconditional
        \STATE Flow matching loss $\mathcal{L}_{\mathrm{FM}}\gets \Vert \bm v- \hat{\bm v}\Vert_2^2$
        \STATE Update $\bm\theta$
        \UNTIL Converged
        \OUTPUT $\bm f_{\bm \phi}$, $\bm f_{\bm \psi}$, $\bm f_{\bm\theta}$, $\mathbf z^{(0)}$, $\texttt{MolEnc}$, $\texttt{AtomEnc}$, $\texttt{LatEnc}$, $\texttt{Atom2Enc}$, $\texttt{MolDec}$, $\texttt{AtomDec}$, $\texttt{PropDec}$
    \end{algorithmic}
\end{algorithm}

\begin{algorithm}[ht]
    \caption{Sampling algorithm of EF-TALFM}\label{algo:sampling}
    \begin{algorithmic}
        \INPUT $\bm f_{\bm \psi}$, $\bm f_{\bm\theta}$, $\texttt{LatEnc}$, $\texttt{Atom2Enc}$, $\texttt{MolDec}$, and $\texttt{AtomDec}$; property $\mathbf y=\emptyset$ if unconditional.
        \STATE Sample $\mathbf z_0\sim\mathcal{N}(\mathbf 0, \mathbf I)$
        \FOR{$t$ in $0, \Delta t, \,2\Delta t, \,\cdots, \,1-\Delta t$}
        \STATE $\mathbf z_{t+\Delta t} \gets \mathbf z_t + \bm f_{\bm\theta}(\mathbf z_t, t, \mathbf y)\,\Delta t$
        \ENDFOR
        \STATE $\hat{\mathbf u} \gets \texttt{MolDec}(\mathbf z_1)$
        \STATE Encode $\bm \nu_0^{(0)} \gets \texttt{LatEnc}(\mathbf z_1)$ 
        \STATE $i\gets 0$
        \REPEAT
        \STATE $i\gets i+1$
        \STATE $\bm \nu_{i-1}^{(L)} \gets f_{\bm \psi}( [\bm \nu_{0}^{(0)}, \cdots ,\bm \nu_{i-1}^{(0)}])$
        \STATE $(\hat{\mathbf x}_i, \hat{\mathbf h}_i)\gets \texttt{AtomDec}(\bm \nu_{i-1}^{(L)})$
        \STATE $\bm \nu_i^{(0)}\gets \texttt{Atom2Enc}(\hat{\mathbf x}_i, \hat{\mathbf h}_i)$
        \UNTIL \texttt{[EOS]} atom-type label is emitted in $\hat{\mathbf h}_i$
        \OUTPUT $(\hat{\mathbf u}, (\hat{\mathbf x}_j,\hat{\mathbf h}_j)_{j=1}^{i-1})$ \hfill $\triangleright$ \textit{Excluding the nonphysical end-of-molecule token at $i$th position}
    \end{algorithmic}
\end{algorithm}

\subsection{Molecular graph reconstruction and geometry refinement} \label{appendix:molecule_reconstruction}
During sampling, EF-TALFM generates atom types, 3D coordinates, atom-level chemical features, and molecule-level attributes, but no bond labels.
Depending on the molecular representation used for training, the decoded sequence contains either only heavy atoms or all atoms, including explicit hydrogens.
We convert either decoded state into a chemically interpretable molecular structure using a fixed, rule-based reconstruction and refinement pipeline that is separate from the learned generator.
The two representations share the same fixed-topology, heavy-atom-restrained refinement principle. 
We illustrate post-decoding molecular graph reconstruction and geometry refinement for a heavy-atom generation example in~\Cref{fig:molecule_graph_reconstruction_and_relaxation}.

\begin{figure}[htbp]
    \centering
    \includegraphics[width=1.0\linewidth]{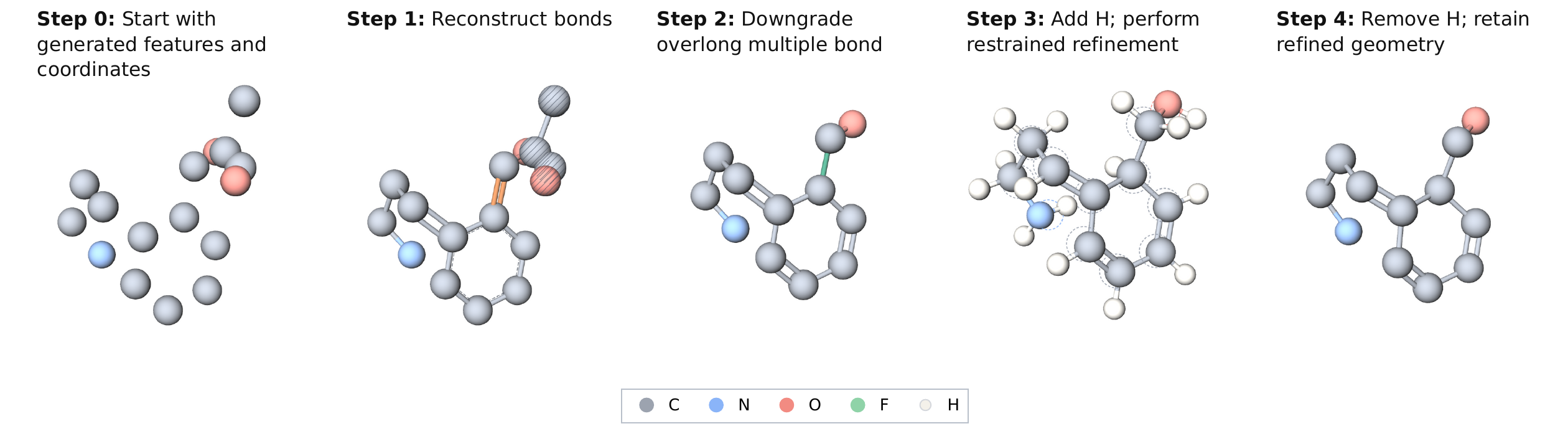}
    \caption{
    \textbf{Deterministic reconstruction and geometry refinement of a generated heavy-atom configuration.} 
    The decoder outputs atom types, 3D coordinates, together with per-atom and molecule-level attributes, but no molecular graph. 
    Connectivity and valence-compatible bond orders are inferred without changing the generated coordinates. 
    The initial graph contains two connected components: the hatched four-heavy-atom propylene oxide fragment is discarded, whereas the 12-heavy-atom component is retained. 
    Brown marks an initially assigned \texttt{C=C} bond whose length is inconsistent with a carbon--carbon double bond. 
    Demotion to \texttt{C--C} (green) dearomatizes the ring. 
    Aromatic bonds before correction are depicted by a solid bond with a thin dashed inner line. 
    Explicit hydrogens are subsequently added for fixed-topology, geometry refinement; dashed element-colored outlines indicate the pre-refinement heavy-atom positions. 
    Finally, the temporary hydrogens are removed while retaining the refined heavy-atom coordinates. 
    }
    \label{fig:molecule_graph_reconstruction_and_relaxation}
\end{figure}

\subsubsection{Molecular graph reconstruction}
Compared to all-atom representations, the heavy-atom only counterpart requires different graph-construction and hydrogen-handling procedures.
In the all-atom setting, the explicit hydrogen positions provide additional geometric evidence for hydrogen parentage and atomic valence, making standard bond perception a strong starting point.
The more demanding reconstruction problem is therefore the heavy-atom setting, in which the hydrogen coordinates are absent, and the graph must be recovered using the decoded heavy-atom geometry and chemical features.

For a heavy-atom decoded sequence, both connectivity and bond orders must be inferred; see Step~1 in \Cref{fig:molecule_graph_reconstruction_and_relaxation}.
Rule-based perception of molecular connectivity and bond orders from three-dimensional coordinates and element identities has been studied previously \citep{zhang2012bondperception}.
Our procedure operates in the same broad setting but additionally uses a distinct set of rules that incorporates the generated chemical attributes.
Specifically, it first infers heavy-atom connectivity from interatomic distances and covalent radii under element-, hybridization-, and ring-aware constraints; then assigns valence-aware bond orders using the generated chemical attributes; and finally applies deterministic valence, aromaticity, and conjugation corrections.
If the resulting graph contains multiple components, we retain one component with the largest number of heavy atoms and discard the others without changing the bonds within the retained component (Step~1).
We then identify double and triple bonds whose interatomic distances exceed element- and bond-order-specific cutoffs based on covalent radii. Their bond orders are lowered one level at a time only when the resulting graph passes RDKit sanitization; no edge is deleted. We refer to this operation as \emph{overlong multiple-bond correction} (Step~2).

\textbf{Heavy-atom molecule reconstruction and geometry refinement.}
To derive a generated heavy-atom molecule's bonds, we first apply a deterministic chemistry-aware reconstruction procedure that is separate from the learned generator. This reconstruction procedure is heuristic rather than guaranteed exact: it maps continuous coordinates and discrete atom-wise predictions into a chemically interpretable heavy-atom molecular graph using chemistry-aware rules rather than learned bond generation.  

Bond topology reconstruction is derived from the atomic coordinates, atom types, and additional chemical descriptors provided by the generator. The descriptors include the atomic number together with auxiliary local chemical attributes such as hybridization class, attached-hydrogen count, formal charge, radical, aromaticity, and ring-membership. We propose candidate heavy-heavy bonds between heavy atom pairs whose Euclidean distance falls below a scaled sum of covalent radii for the pair. Candidate edges are then greedily accepted in order of increasing distance, subject to element- and hybridization-dependent valence limits and soft preferred-degree constraints.

Bond orders are assigned in a second deterministic stage. Starting from the inferred heavy-atom connectivity graph, all accepted edges are initialized as single bonds and are then selectively promoted to double or triple bonds according to valence deficits implied by the predicted atom-wise features. The procedure further applies a small number of chemistry-motivated corrections, including Kekul\'e assignment for aromatic-like six-membered rings, strict valence clipping, and terminal oxo promotion when valence permits. The resulting heavy-atom graph is converted into an RDKit molecule, and optional per-atom formal-charge, radical, aromaticity, conjugation, and ring-membership annotations are applied through deterministic post-processing. A complete reconstruction algorithm is given in \Cref{algo:reconstruction}.

Next we augment the existing heavy-atom reconstruction pipeline with a lightweight chemistry-aware geometry refinement stage designed to improve structural validity.
Starting from the initially reconstructed heavy-atom graph and the bond orders inferred from atom- and molecule-level features, we retain the largest connected component and discard the rest, since the reconstructed graph may contain multiple components. 

We first apply overlong multiple-bond correction, selectively downgrading a multiple bond when its interatomic distance is too long for the assigned bond order. We then add hydrogens and perform restrained force-field relaxation, softly constraining heavy atoms to remain close to the predicted scaffold while allowing local geometric corrections.
This fixed post-processing pipeline improves geometric plausibility while preserving the model's original 3D prediction as much as possible.

\textbf{All-atom molecule reconstruction and geometry refinement.}
For an all-atom decoded sequence, we use the rule-based bond-perception routines in Open Babel~\citep{oboyle2011openbabel}, which take generated atom types and 3D coordinates and produce an initial candidate graph comprising heavy--heavy and heavy--hydrogen connectivity and candidate heavy--heavy bond orders.
We inspect this raw graph for connectivity and the number of explicit hydrogen neighbors of each heavy atom against the decoder's attached-hydrogen feature. 
If either check fails, we conditionally reconstruct the hydrogen layer by removing the explicit hydrogens, while retaining Open Babel's heavy-atom connectivity, bond orders, formal charges, and radical states, and rematerializing the predicted number of hydrogens on each heavy atom. 
This operation can repair missing or incorrectly attached hydrogens, but does not reconstruct or reconnect the heavy-atom graph. 
We then apply the same connectivity-preserving overlong multiple-bond correction as in the heavy-atom branch.

For geometry refinement, topology and explicit composition are retained throughout relaxation.
Under the canonical ordering used for this representation, most ordinary hydrogens occur after the heavy atoms and are therefore among the later autoregressive predictions.
Before the joint restrained relaxation, we replace the coordinate of an ordinary trailing hydrogen with a valence-based initialization only when its bond to the assigned parent has an implausible length or when it has a severe nonbonded clash.  
We first optimize the ordinary trailing hydrogens with all other atoms fixed. We then optimize the complete molecule while restraining the heavy-atom scaffold and leaving hydrogens unrestrained.
Thus, as in the heavy-atom branch, it performs inexpensive local refinement without changing the selected molecular topology. However, unlike the heavy-atom branch, which returns an implicit-hydrogen molecule, the all-atom branch returns an explicit-hydrogen molecule.

\subsubsection{Geometry refinement}
Short force-field minimization has been used to refine the internal geometries of generated molecules \citep{ragoza2022generating}, and recent benchmarking shows that local relaxation can substantially improve the conformational validity of generated 3D molecules \citep{baillif2024genbench3d}.
For heavy-atom generation, RDKit first adds temporary explicit hydrogens; all heavy atoms are then restrained near their decoded positions during minimization (Step~3), and the refined heavy-atom coordinates are transferred back while the temporary hydrogens are discarded (Step~4).

\begin{algorithm}[htbp]
    \caption{Heavy-atom graph reconstruction from generated coordinates and atom-wise features}\label{algo:reconstruction}
    \begin{algorithmic}[1]
        \INPUT Heavy-atom coordinates $\hat{\mathbf R} = (\hat{\mathbf r}_1,\dots,\hat{\mathbf r}_N)\in\mathbb{R}^{N\times 3}$, atomic numbers $\hat{\mathbf z}$, hybridization tags $\hat{\mathbf h}^{\mathrm{hyb}}$, chiral tags $\hat{\mathbf h}^{\mathrm{chiral}}$, attached-hydrogen counts $\hat{\mathbf n}^{\mathrm H}$, predicted total charge $\hat Q$, predicted spin multiplicity $\hat s$, optional formal-charge tags $\hat{\mathbf q}$, optional radical tags $\hat{\mathbf r}^{\mathrm{rad}}$, optional atom-level conjugation/aromaticity/ring hints, and optional molecule-level ring descriptors
        \STATE Decode per-atom formal charges from $\hat{\mathbf q}$ when available
        \STATE Select reconstruction policy $\pi \gets \texttt{ChoosePolicy}(\hat{\mathbf z}, \hat{\mathbf R}, \hat Q)$
        \STATE Initialize candidate bond list $\mathcal C \gets \emptyset$
        \FOR{$1 \le i < j \le N$}
            \STATE Compute distance $d_{ij} \gets \|\hat{\mathbf r}_i - \hat{\mathbf r}_j\|_2$
            \STATE Compute bond threshold $\tau_{ij} \gets \lambda_\pi \big(r^{\mathrm{cov}}(\hat z_i) + r^{\mathrm{cov}}(\hat z_j)\big)$
            \IF{$d_{ij} \le \tau_{ij}$}
                \STATE Append $(d_{ij}, i, j)$ to $\mathcal C$
            \ENDIF
        \ENDFOR
        \STATE Sort $\mathcal C$ by increasing distance
        \STATE Compute per-atom maximum valence targets $\{v_i^{\max}\}_{i=1}^N$
        \STATE Compute per-atom preferred heavy-atom degrees $\{d_i^\star\}_{i=1}^N$
        \STATE Initialize heavy-atom bond set $E \gets \emptyset$ and degrees $\deg(i)\gets 0$
        \FORALL{$(d_{ij}, i, j)\in \mathcal C$ in sorted order}
            \IF{$\deg(i) < v_i^{\max}$ and $\deg(j) < v_j^{\max}$}
                \IF{adding $(i,j)$ does not violate the preferred-degree constraint}
                    \IF{optional ring-count / fused-ring constraints are satisfied}
                        \STATE Add $(i,j)$ to $E$
                        \STATE Update $\deg(i)$ and $\deg(j)$
                    \ENDIF
                \ENDIF
            \ENDIF
        \ENDFOR
        \STATE Initialize all bond orders as single: $b_{ij}\gets 1$ for all $(i,j)\in E$
        \STATE Compute target atom valences $\{t_i\}_{i=1}^N$ from $\hat{\mathbf z}$, $\hat{\mathbf h}^{\mathrm{hyb}}$, $\hat{\mathbf n}^{\mathrm H}$, optional decoded formal charges, and optional radical tags
        \STATE Compute valence deficits $\delta_i \gets t_i - (\deg(i)+\hat n_i^{\mathrm H})$
        \REPEAT
            \STATE Score each bond $(i,j)\in E$ for possible bond-order promotion using current deficits and local hybridization constraints
            \STATE Let $(i^\star,j^\star)$ be the highest-scoring promotable bond
            \IF{no promotable bond remains}
                \STATE \textbf{break}
            \ENDIF
            \STATE Increase bond order $b_{i^\star j^\star}\gets b_{i^\star j^\star}+1$
            \STATE Update deficits $\delta_{i^\star}, \delta_{j^\star}$
        \UNTIL No promotable bond remains
        \STATE Apply aromatic-like six-ring Kekul\'e correction to $\{b_{ij}\}$
        \STATE Clip bond orders to satisfy stricter valence limits
        \STATE Promote eligible terminal oxo bonds from single to double order
        \STATE Build heavy-atom RDKit molecule
        $\hat G_{\mathrm{heavy}} \gets \texttt{BuildMol}(\hat{\mathbf z}, \hat{\mathbf R}, E, \{b_{ij}\}, \hat{\mathbf h}^{\mathrm{chiral}}, \hat{\mathbf q}, \hat{\mathbf r}^{\mathrm{rad}})$
        \IF{atom-level conjugation / aromaticity / ring hints are available}
            \STATE $\hat G_{\mathrm{heavy}} \gets \texttt{ApplyAtomFeatureHints}(\hat G_{\mathrm{heavy}})$
        \ENDIF
        \STATE Compute reconstructed total formal charge from $\hat G_{\mathrm{heavy}}$
        \STATE Compute multiplicity-parity consistency from input atomic numbers, attached-hydrogen counts, total charge $\hat Q$, and spin multiplicity $\hat s$
        \STATE Compute inferred ring count and inferred fused-ring indicator from $E$
        \OUTPUT Reconstructed heavy-atom molecule $\hat G_{\mathrm{heavy}}$, bond set $E$, bond orders $\{b_{ij}\}$, and consistency diagnostics
    \end{algorithmic}
\end{algorithm}

\FloatBarrier

\textbf{Refinement parameters and optimization settings.} Hydrogen-inclusive force-field relaxation uses a flat-bottom positional restraint on every heavy atom, with an unpenalized displacement radius of $0.15\,\text{\AA}$, a restraint force constant of $200\,\mathrm{kcal\,mol^{-1}\,\text{\AA}^{-2}}$, and at most $200$ minimization iterations. The $0.15\,\text{\AA}$ radius allows small local corrections while remaining much smaller than a typical heavy-atom bond length; the force constant strongly penalizes larger deviations; and the iteration budget is intended for inexpensive local cleanup rather than conformational search. These fixed values are chemistry-motivated practical defaults rather than formally optimized hyperparameters. Although force-field refinement is conventional in coordinate-based 3D molecule-generation pipelines, prior work does not establish these exact restraint values; we therefore evaluate the effect of the refinement separately from graph reconstruction.

\subsection{Canonical atom position alignment}\label{appendix:atom_position_alignment}
We define a molecule's canonical pose relative to a fixed canonical atom ordering. 
Once atom rows are placed in a deterministic order, a deterministic rigid-frame construction can select one representative from the molecule's \(\mathrm{SE}(3)\) orbit.
This lets a non-equivariant model operate on coordinates in a consistent canonical frame.
By construction, two coordinate inputs $\mathbf{x}$ and $\mathbf{x}'$ that are related by an element of $\mathrm{SE}(3)$ and share the same canonical atom ordering produce the same aligned coordinates $\mathbf{x}^{\mathrm{final}}$, up to numerical precision. Thus, the downstream model need not learn invariance to translations or proper rotations from data.

The alignment first translates the first canonical atom to the origin. It then selects the first later canonical atom and sets its target position to $(\|\mathbf{u}\|, 0, 0)$, i.e., on the positive $x$-axis at its true distance from the origin.
To fix the remaining proper-rotation degree of freedom, the algorithm searches in canonical order for the first atom that is sufficiently non-collinear with this axis, as specified in \Cref{algo:alignment}.
The selected anchors determine a proper rotation into the standardized coordinate frame.

If no sufficiently non-collinear anchor exists, the molecule is treated as collinear for alignment purposes. 
All atoms are placed on the $x$ axis by signed projection onto the canonical axis, preserving their signed axial coordinates.
For an exactly collinear geometry, this preserves all pairwise distances and avoids an arbitrary rotation around the molecular axis.
For a nearly collinear geometry, it removes transverse components, thus the projection can introduce small geometric changes and erase small three-dimensional signed volumes. Such an output is accepted only if it passes the final distance check. If no axis anchor exists, all coordinates are set to zero; a single atom is translated to the origin.

When a non-collinear anchor is found, the resulting proper rotation preserves geometric handedness and keeps mirror-image non-planar conformations distinct. 
Reflected exactly planar or collinear point sets can already be related by a proper rotation. 
The nearly collinear projection branch is an approximation and does not provide the same handedness guarantee.

For PCQM4Mv2 preprocessing, we use the direct basis solve and keep the first atom at the origin.
The algorithm takes canonically ordered coordinates as input and uses a single numerical tolerance, $\mathtt{tol}$, for anchor selection, the internal numerical checks, and the final pairwise-distance check.
All coordinate values are expressed numerically in \AA; the fixed tolerance and complete procedure are given in \Cref{algo:alignment}.

\begin{algorithm}[ht]
   \caption{Atom position alignment}
    \begin{algorithmic}
    \INPUT Canonically ordered positions of $n\ge1$ atoms $\mathbf{x}\in\mathbb{R}^{n\times 3}$, expressed numerically in \AA.
    \STATE \textbf{Set tolerance:} $\mathtt{tol}\gets10^{-4}$. \COMMENT{Allow numerical roundoff while keeping distance errors small relative to bond lengths.}
    \STATE \textbf{Translate:}
    $\mathbf{x}^{\mathrm{trans}}\gets \mathbf{x}-\mathbf{x}_1$.
    \IF{$n=1$}
        \STATE $\mathbf{x}^{\mathrm{final}}\gets \mathbf{x}^{\mathrm{trans}}$
    \ELSE
    \STATE Find the first index $j \gets \min\{i\in\{2,\ldots,n\}: \|\mathbf{x}^{\mathrm{trans}}_i\|>0\}$.

    \IF{no such $j$ exists}
        \STATE  $\mathbf{x}^{\mathrm{final}}\gets\mathbf{0}\in\mathbb{R}^{n\times 3}$
    \ELSE
        \STATE Let $\mathbf{u}\gets \mathbf{x}^{\mathrm{trans}}_j$.
        \STATE Find the first index $k \gets \min\left\{
              i\in\{2,\ldots,n\}\setminus\{j\}:
              \|\mathbf{u}\times \mathbf{x}^{\mathrm{trans}}_i\|
              >
              \mathtt{tol}\max\!\left(\|\mathbf{u}\|\|\mathbf{x}^{\mathrm{trans}}_i\|,1\right)
              \right\}$. \hfill \COMMENT{\textit{Non-collinearity.}}

            \IF{no such $k$ exists}
                \STATE \textbf{Collinear case:} set
                $\hat{\mathbf{u}}\gets \mathbf{u}/\|\mathbf{u}\|$.
                \FOR{$i=1,\ldots,n$}
                     \STATE Set $\mathbf{x}^{\mathrm{final}}_i \gets
                     \left(\left\langle \mathbf{x}^{\mathrm{trans}}_i,\hat{\mathbf{u}}\right\rangle,0,0\right)$.
                \ENDFOR
            \ELSE
                \STATE \textbf{Initialize target frame:}
                $\mathbf{x}^{\mathrm{temp}}\gets \mathbf{0}\in\mathbb{R}^{n\times 3}$.
                \STATE Compute
                 \[
                    d_{1j}\gets \|\mathbf{x}^{\mathrm{trans}}_j-\mathbf{x}^{\mathrm{trans}}_1\|,
                    \quad
                    d_{1k}\gets \|\mathbf{x}^{\mathrm{trans}}_k-\mathbf{x}^{\mathrm{trans}}_1\|,
                    \quad
                    d_{jk}\gets \|\mathbf{x}^{\mathrm{trans}}_k-\mathbf{x}^{\mathrm{trans}}_j\|.
                \]
                \STATE Set
                \[
                  \mathbf{x}^{\mathrm{temp}}_j\gets (d_{1j},0,0).
                \]
                \STATE Compute
                \[
                  r_{k1}\gets \frac{d_{1j}^2-d_{jk}^2+d_{1k}^2}{2d_{1j}}.
                \]
                \STATE Set
                \[
                  \mathbf{x}^{\mathrm{temp}}_k
                  \gets
                  \left(
                  r_{k1},
                  0,
                  \sqrt{\max\{d_{1k}^2-r_{k1}^2,0\}}
                  \right).
                \]

                \STATE \textbf{Construct column bases:}
                \[
                \begin{aligned}
                \mathbf A&\gets[\mathbf{x}^{\mathrm{trans}}_j,\mathbf{x}^{\mathrm{trans}}_k,
                 \mathbf{x}^{\mathrm{trans}}_j\times\mathbf{x}^{\mathrm{trans}}_k],\\
                \mathbf B&\gets[\mathbf{x}^{\mathrm{temp}}_j,\mathbf{x}^{\mathrm{temp}}_k,
                 \mathbf{x}^{\mathrm{temp}}_j\times\mathbf{x}^{\mathrm{temp}}_k].
                \end{aligned}
                \]
                \STATE \textbf{Solve:} $\mathbf R\mathbf B=\mathbf A$. Reject a negative determinant. \hfill \COMMENT{\textit{Exclude orientation reversal to preserve handedness.}}
                \STATE Let $\mathbf X_b=\mathbf x^{\mathrm{trans}}_{[1,j,k]}$ and $\mathbf Y_b=\mathbf x^{\mathrm{temp}}_{[1,j,k]}$ contain the three anchor rows.
                \STATE Compute $\mathbf t_b=\operatorname{mean}(\mathbf Y_b)-\operatorname{mean}(\mathbf X_b)\mathbf R$.
                \STATE Reject if both $\|\mathbf X_b\mathbf R+\mathbf 1_3\mathbf t_b-\mathbf Y_b\|_{\max}>\mathtt{tol}$ and $\|\mathbf R\mathbf R^\top-\mathbf I_3\|_{\max}>\mathtt{tol}$. \hfill \COMMENT{$\|\cdot\|_{\max}$: \textit{the largest absolute matrix entry.}}
                \STATE \textbf{Rotate:}
                $\mathbf{x}^{\mathrm{final}}\gets \mathbf{x}^{\mathrm{trans}}\mathbf{R}$.
                \ENDIF
                \ENDIF
                \ENDIF
                \STATE \textbf{Check distances for $n\ge2$:} reject unless $\max_{1\le i<\ell\le n}
                \left|
                \|\mathbf{x}^{\mathrm{final}}_i-\mathbf{x}^{\mathrm{final}}_\ell\|
                -
                \|\mathbf{x}_i-\mathbf{x}_\ell\|
                \right|
                \le \mathtt{tol}$.
                \OUTPUT Canonically aligned positions $\mathbf{x}^{\mathrm{final}}$.
   \end{algorithmic}
   \label{algo:alignment}
\end{algorithm}

Canonical SMILES fixes a graph-level representation independently of the input atom indices.
For graphs with nontrivial automorphisms, however, the back-matching step is not uniquely determined and may tie-break using the input ordering; we therefore do not claim a mathematically unique atom labeling for all symmetric molecules, only a deterministic ordering for a fixed input that is used consistently across training and evaluation.

\subsection{Chemical state variables} \label{appendix:additional_chemical_features}
Molecular geometry is represented as atom types and their 3D coordinates, without explicitly generating pairwise bonds. Instead, atom-wise features provide implicit context about bonding, stereochemistry, and charge, allowing the model to place atoms in space in a way that is consistent with the underlying molecular structure. Each of these atom-wise features in $\mathbf h$ encodes constraints that would otherwise be enforced through explicit bonds, giving the model enough local chemical context to place atoms consistently in 3D:

\begin{itemize}
    \item \textbf{Atomic number} determines elemental identity and typical valence, atomic size, and preferred bonding environments, thereby setting the baseline for how many neighbors an atom can support and at what distances.
    \item \textbf{Chirality tag} specifies stereochemical configuration (e.g., tetrahedral centers), guiding the model toward correct 3D arrangements (e.g., avoiding mirror inversions at chiral centers).
    \item \textbf{Formal charge} constrains electron count and influences preferred bonding patterns and geometry (e.g., planar vs. pyramidal nitrogen).
    \item \textbf{Number of attached hydrogens} helps fix valence and saturates local bonding, reducing ambiguity in how many heavy-atom neighbors must be placed nearby.
    \item \textbf{Hybridization tag} encodes orbital geometry (sp, sp², sp³), directly informing expected bond angles and local spatial arrangement (linear, trigonal planar, tetrahedral).
    \item \textbf{Number of radical electrons} identifies open-shell local electronic states and constrains valence assignments.
    \item \textbf{Aromaticity tag} signals delocalized $\pi$-systems, encouraging planar ring structures and characteristic bond-length patterns.
    \item \textbf{Conjugation tag} marks atoms participating in conjugated bonding environments.
    \item \textbf{Is-in-ring tag} indicates topological constraints that restrict geometry (e.g., ring closure, limited flexibility, characteristic angles/strain).
\end{itemize}

These features are provided by RDKit for each molecule and are encoded into categorical integer labels based on the convention of the Open Graph Benchmark (OGB) \citep{hu2021ogblsc}. For additional chemical context for generation, we include several molecule-wide descriptors:

\begin{itemize}
    \item \textbf{Total charge} specifies the net molecular charge, constraining the overall electron count and ensuring consistency with atom-level formal charges.
    \item \textbf{Spin multiplicity} encodes the global electronic state (e.g., singlet, doublet), determining the number of unpaired electrons and restricting allowable radical configurations.
    \item \textbf{Number of rings} provides a coarse topological constraint on cyclic structure, helping guide ring formation and overall connectivity.
    \item \textbf{Fused ring indicator} specifies whether the molecule contains fused ring systems, constraining how rings can share atoms and influencing feasible geometries and bonding patterns.
\end{itemize}

\subsection{Latent regularization}\label{appendix:latent_regularization}
Let $\boldsymbol{\mu}^{(b)} \in \mathbb{R}^d$ denote the posterior mean of sample $b$ in a minibatch of size $B$, and let
$\mu_j^{(b)}$ be its $j$-th latent coordinate. We define
\begin{equation}
\mathcal{L}_{\mathrm{var}}
=
\frac{1}{d}\sum_{j=1}^{d}
\max\!\left(0,\tau - \operatorname{Var}_{b}\!\big[\mu_j^{(b)}\big]\right),
\qquad \tau = 0.01,
\label{eq:vae_var_floor}
\end{equation}
which softly enforces a minimum variance floor on each latent mean dimension, and
\begin{equation}
\mathcal{L}_{\mathrm{cov}}
=
\frac{1}{d}
\left\|
\operatorname{offdiag}\!\big(\operatorname{Cov}_{b}(\boldsymbol{\mu}^{(b)})\big)
\right\|_{F}^{2},
\label{eq:vae_cov_penalty}
\end{equation}
Together, the variance floor discourages inactive latent coordinates, while the covariance penalty discourages redundant coordinates in the minibatch posterior means.

\subsection{Learning and sampling from posterior mean}\label{appendix:learning_and_sampling_from_posterior_mean}
Unlike latent generative models that sample from the full aggregated VAE posterior, we construct the flow-training targets from the posterior means $\mathbf z_\mu$. Before flow training, a fitted invertible affine map $\mathcal W$ whitens these means; the flow operates on $\widetilde{\mathbf z}_\mu=\mathcal W(\mathbf z_\mu)$, and $\mathcal W^{-1}$ maps generated terminal states back to the decoder's latent coordinates.
Using posterior means removes stochastic posterior noise from the target distribution and gives the flow model a deterministic representation of each molecule.
The posterior scales $\mathbf z_\sigma$ are still important during autoencoder training: together with the KL penalty, they regularize the latent geometry and encourage the decoder to remain stable under local perturbations. 
Thus, the flow model learns the distribution of molecule-level latent codes, while the VAE regularization shapes this latent space to be smoother and more suitable for generation.

\clearpage

\section{Training configurations}\label{appendix:hyperparameters}

\subsection{UAE-3D: unconditional generation}\label{appendix:hyperparameter_uae_3d}
\begin{table}[H]
\centering 
\caption{Hyperparameter setup for PCQM4Mv2 unconditional VAE training in UAE-3D.}
\label{tab:pcqm4mv2-vae-hparams}
\begin{tabular}{@{}p{0.20\textwidth}p{0.52\textwidth}p{0.20\textwidth}@{}}
\toprule
Category & Hyperparameter & Value \\
\midrule
Data & Batch size & $512$ \\
Data & Data loader workers & $8$ \\
Data & Rotation augmentation & \texttt{enabled} \\
Data & Translation augmentation & \texttt{enabled} \\
Data & Translation scale & $0.1$ \\
Data & Center atomic coordinates & \texttt{enabled} \\
Data & Position standard deviation & \texttt{estimated from $10{,}000$ training samples} \\
Data & Training/validation batch fraction & $1.0$ / $1.0$ \\
Optimization & Learning rate & $5 \times 10^{-5}$ \\
Optimization & Weight decay & $1 \times 10^{-5}$ \\
Optimization & Gradient accumulation & $1$ \\
Optimization & Gradient clipping & $1.0$ \\
Optimization & Precision & \texttt{bf16-mixed} \\
Training & Max epochs & $2{,}000$ \\
Training & Check validation every $n$ epochs & $5$ \\
Training & Save checkpoint every $n$ epochs & $20$ \\
Training & Cache best-validation checkpoint every $n$ epochs & $5$ \\
Training & Test every $n$ epochs & $200$ \\
Architecture & Encoder hidden dimension & $64$ \\
Architecture & Encoder attention heads & $8$ \\
Architecture & Encoder blocks & $6$ \\
Architecture & Latent dimension & $4$ \\
Architecture & Decoder hidden dimension & $64$ \\
Architecture & Decoder attention heads & $8$ \\
Architecture & Decoder blocks & $4$ \\
Architecture & Dropout & $0.1$ \\
Loss & Atom loss weight & $1.0$ \\
Loss & Bond loss weight & $1.0$ \\
Loss & Coordinate loss weight & $1.0$ \\
Loss & Distance loss weight & $1.0$ \\
Loss & Bond distance loss weight & $10.0$ \\
Loss & KLD weight & $1 \times 10^{-8}$ \\
Loss & Center prediction & \texttt{disabled} \\
Loss & Alignment prediction & \texttt{disabled} \\
\bottomrule
\end{tabular}
\end{table} 

\begin{table}[H]
\centering 
\caption{Hyperparameter setup for PCQM4Mv2 unconditional latent diffusion model (LDM) training in UAE-3D.}
\label{tab:pcqm4mv2-ldm-hparams}
\begin{tabular}{@{}p{0.20\textwidth}p{0.48\textwidth}p{0.25\textwidth}@{}}
\toprule
Category & Hyperparameter & Value \\
\midrule
Data & Batch size & $512$ \\
Data & Data loader workers & $8$ \\
Data & Number of generated samples for test-time sampling & $10{,}000$ \\
Data & Rotation augmentation & \texttt{enabled} \\
Data & Translation augmentation & \texttt{enabled} \\
Data & Translation scale & $0.1$ \\
Data & Center atomic coordinates & \texttt{enabled} \\
Data & Position standard deviation & \texttt{estimated from $10{,}000$ training samples} \\
Data & Training/validation batch fraction & $1.0$ / $1.0$ \\
Optimization & Learning rate & $1 \times 10^{-4}$ \\
Optimization & Weight decay & $0.05$ \\
Optimization & Minimum learning rate & $1 \times 10^{-5}$ \\
Optimization & Warmup learning rate & $1 \times 10^{-6}$ \\
Optimization & Warmup steps & $1{,}000$ \\
Optimization & Learning-rate scheduler & \texttt{linear\_warmup\_cosine} \\
Optimization & Gradient accumulation & $1$ \\
Optimization & Gradient clipping & $1.0$ \\
Optimization & Precision & \texttt{bf16-mixed} \\
Training & Max epochs & $10{,}000$ \\
Training & Check validation every $n$ epochs & $5$ \\
Training & Save checkpoint every $n$ epochs & $5$ \\
Training & Cache best-validation checkpoint every $n$ epochs & $5$ \\
Training & Test every $n$ epochs & $200$ \\
VAE backbone & Encoder hidden dimension & $64$ \\
VAE backbone & Encoder attention heads & $8$ \\
VAE backbone & Encoder blocks & $6$ \\
VAE backbone & Latent dimension & $4$ \\
VAE backbone & Decoder hidden dimension & $64$ \\
VAE backbone & Decoder attention heads & $8$ \\
VAE backbone & Decoder blocks & $4$ \\
VAE backbone & Dropout & $0.1$ \\
Objective & Atom loss weight & $1.0$ \\
Objective & Bond loss weight & $1.0$ \\
Objective & Coordinate loss weight & $1.0$ \\
Objective & Distance loss weight & $1.0$ \\
Objective & Bond distance loss weight & $10.0$ \\
Objective & KLD weight & $1 \times 10^{-8}$ \\
Objective & Center prediction & \texttt{disabled} \\
Objective & Alignment prediction & \texttt{disabled} \\
Diffusion & Backbone & \texttt{DiT enabled} \\
Diffusion & Conditioning & \texttt{none (unconditional)} \\
Diffusion & Hidden dimension & $512$ \\
Diffusion & Attention heads & $8$ \\
Diffusion & Layers & $8$ \\
Diffusion & MLP ratio & $4.0$ \\
Diffusion & Dropout & $0.0$ \\
Diffusion & Latent whitening & \texttt{isotropic} \\
Diffusion & DDPM denoising steps & $100$ \\
Diffusion & Noise temperature & $0.95$ \\
Diffusion & Classifier-free guidance drop probability & $0.1$ \\
Diffusion & Classifier-free guidance weight & $0.5$ \\
Noise schedule & Scheduler & \texttt{cosine} \\
Noise schedule & Continuous $\beta_0$ & $0.1$ \\
Noise schedule & Continuous $\beta_1$ & $20$ \\
\bottomrule
\end{tabular}
\end{table}

\subsection{FlowMol: unconditional generation}\label{appendix:hyperparameter_flowmol}

\begin{table}[H]
\centering 
\renewcommand{\arraystretch}{1.12}
\caption{Hyperparameter setup for PCQM4Mv2 unconditional FlowMol training.}
\label{tab:pcqm4mv2-flowmol-hparams}
\begin{tabular}{@{}p{0.25\textwidth}p{0.4\textwidth}p{0.25\textwidth}@{}}
\toprule
Category & Hyperparameter & Value \\
\midrule
Data & \texttt{atom\_map} & 21 PCQM4Mv2 atom types \\
 & \texttt{explicit\_aromaticity} & \texttt{true} \\
 & \texttt{batch\_size} & $16$ \\
 & \texttt{num\_workers} & $6$ \\
 & \texttt{max\_num\_edges} & $80{,}000$ \\ 
 & \texttt{precision} & \texttt{bf16-mixed} \\
 & \texttt{max\_epochs} & $18$ \\
 & \texttt{accumulate\_grad\_batches} & $5$ \\
 & \texttt{val\_loss\_interval} & $0.25$ \\ 
Optimization & \texttt{base\_lr} & $1 \times 10^{-4}$ \\
 & \texttt{warmup\_length} & $0.044$ \\
 & \texttt{weight\_decay} & $1 \times 10^{-12}$ \\
 & \texttt{restart\_interval} / \texttt{restart\_type} & $0$ / \texttt{linear} \\
 & \texttt{ema\_decay} & $0$ \\
Flow matching & \texttt{parameterization} & \texttt{ctmc} \\
 & \texttt{time\_scaled\_loss} & \texttt{true} \\
 & \texttt{total\_loss\_weights} of $(x,a,c,e)$ & $(3.0,0.4,1.0,2.0)$ \\
 & \texttt{distort\_p} & $0.2$ \\
 & \texttt{fake\_atom\_p} & $0.0$ \\
 & \texttt{n\_atom\_charges} & $6$ \\
Priors and schedules & \texttt{x} prior & centered normal, std. $1.0$, aligned \\
 & \texttt{a} / \texttt{c} / \texttt{e} priors & \texttt{ctmc}, unaligned \\
 & \texttt{schedule\_type} & linear for $x$, $a$, $c$, and $e$ \\
Vector field & \texttt{self\_conditioning} & \texttt{true} \\
 & \texttt{n\_hidden\_scalars} & $512$ \\
 & \texttt{n\_hidden\_edge\_feats} & $256$ \\
 & \texttt{n\_vec\_channels} & $64$ \\
 & \texttt{n\_molecule\_updates} & $8$ \\
 & \texttt{n\_message\_gvps} / \texttt{n\_update\_gvps} & $3$ / $3$ \\
 & \texttt{n\_expansion\_gvps} & $3$ \\
 & \texttt{n\_recycles} / \texttt{convs\_per\_update} & $1$ / $1$ \\
 & \texttt{n\_cp\_feats} & $4$ \\
 & \texttt{update\_edge\_w\_distance} & \texttt{true} \\
 & \texttt{message\_norm} & \texttt{sum} \\
 & \texttt{rbf\_dmax} / \texttt{rbf\_dim} & $10$ / $32$ \\
 & \texttt{time\_embedding\_dim} & $64$ \\
 & \texttt{a\_token\_dim, c\_token\_dim, e\_token\_dim} & $64$ each \\
 & \texttt{stochasticity / high\_confidence\_threshold} & $30.0$ / $0.9$ \\
 & \texttt{dropout} & $0.1$ \\
\bottomrule
\end{tabular}
\end{table}

\subsection{EF-TALFM: unconditional generation}\label{appendix:hyperparameter_talfm}
\begin{table}[H]
\centering 
\renewcommand{\arraystretch}{1.12}
\caption{Hyperparameters for PCQM4Mv2 unconditional EF-TAVAE training.}
\label{tab:deltaai-vae-task7}
\begin{tabular}{@{}p{0.25\textwidth}p{0.4\textwidth}p{0.25\textwidth}@{}}
\toprule
Category & Hyperparameter & Value \\ 
\midrule
Model architecture & \texttt{model} & \texttt{transformer-p} \\
 & \texttt{norm\_first} & \texttt{true} \\
 & \texttt{latent\_dim} & $32$ \\
 & \texttt{encoder\_embed\_dim} & $384$ \\
 & \texttt{encoder\_n\_layers} & $6$ \\
 & \texttt{encoder\_n\_heads} & $12$ \\
 & \texttt{encoder\_dim\_feedforward} & $768$ \\
 & \texttt{encoder\_width\_multiplier} & $3.0$ \\
 & \texttt{encoder\_depth\_multiplier} & $3.0$ \\
 & \texttt{decoder\_embed\_dim} & $384$ \\
 & \texttt{decoder\_n\_layers} & $6$ \\
 & \texttt{decoder\_n\_heads} & $12$ \\
 & \texttt{decoder\_dim\_feedforward} & $1{,}536$ \\
 & \texttt{decoder\_width\_multiplier} & $3.0$ \\
 & \texttt{decoder\_depth\_multiplier} & $3.0$ \\
\midrule
Optimization / training & \texttt{precision} & \texttt{bf16-mixed} \\
 & \texttt{lr} & $1\times10^{-3}$ \\
 & \texttt{lr\_end} & $1\times10^{-6}$ \\
 & \texttt{lr\_scheduler} & \texttt{cosine\_with\_lr\_end} \\
 & \texttt{warmup\_proportion} & $0.0$ \\
 & \texttt{max\_steps} & $400{,}000$ \\
 & \texttt{num\_decay\_steps} & $-1$ \\
 & \texttt{batch\_size} & $512$ \\
 & \texttt{accumulate\_grad\_batches} & $1$ \\
 & \texttt{optimizer} & \texttt{adamw} \\
 & \texttt{adam\_betas} & $(0.9, 0.98)$ \\
 & \texttt{adam\_eps} & $10^{-8}$ \\
 & \texttt{weight\_decay} & $0.01$ \\
 & \texttt{gradient\_clip\_val} & $5$ \\
\midrule
Loss / regularization & \texttt{pos\_loss\_type} & \texttt{smoothl1-0.001} \\
 & \texttt{position\_scale} & $1.0$ \\
 & \texttt{pos\_weight} & $10$ \\
 & \texttt{seq\_weight} & $1$ \\
 & \texttt{charge\_weight} & $1$ \\
 & \texttt{spinmp\_weight} & $1$ \\
 & \texttt{kl\_weight} & $1\times10^{-3}$ \\
 & \texttt{vae\_noise\_scale} & $0$ \\
 & \texttt{vae\_disable\_learned\_variance} & \texttt{false} \\
 & \texttt{vae\_corruption\_ratio} & $0.1$ \\
 & \texttt{vae\_free\_bits\_per\_dim} & $0.02$ \\
 & \texttt{vae\_sigma\_clamp\_min} & $0.01$ \\
 & \texttt{vae\_sigma\_clamp\_max} & $1.5$ \\
 & \texttt{vae\_mu\_var\_floor} & $0.01$ \\
 & \texttt{vae\_mu\_var\_floor\_weight} & $5\times10^{-3}$ \\
 & \texttt{vae\_mu\_cov\_weight} & $5\times10^{-4}$ \\
 & \texttt{dropout} & $0.0$ \\
 & \texttt{max\_positions} & $64$ \\
\bottomrule
\end{tabular}
\end{table}
In the reported EF-TAVAE configuration, posterior scales are clamped to $[0.01,1.5]$, the KL term uses $0.02$ free nats per latent dimension, and 10\% of teacher-forced atom-type inputs are masked. The coordinate, categorical-sequence, charge, and spin-multiplicity loss weights are $10$, $1$, $1$, and $1$, respectively; $\beta_{\mathrm{KL}}=10^{-3}$, $\lambda_{\mathrm{var}}=5\times10^{-3}$, and $\lambda_{\mathrm{cov}}=5\times10^{-4}$. These settings correspond to the configuration entries in the table above.
\begin{table}[H]
\centering 
\renewcommand{\arraystretch}{1.12}
\caption{Hyperparameter setup for PCQM4Mv2 unconditional latent flow matching training in EF-TALFM.}
\label{tab:deltaai-lfm-task310}
\begin{tabular}{@{}p{0.25\textwidth}p{0.4\textwidth}p{0.25\textwidth}@{}}
\toprule
Category & Hyperparameter & Value \\
\midrule
Pretrained VAE & \texttt{pretrained\_vae\_subfolder} & \texttt{vae} \\
 & \texttt{pretrained\_vae\_checkpoint} & \texttt{last.ckpt} \\
\midrule
Latent data setup & \texttt{model} & \texttt{transformer-p} \\
 & \texttt{probabilistic\_model} & \texttt{flow\_matching} \\
 & \texttt{context\_dim} & $0$ \\
 & \texttt{gm\_use\_latent\_stats} & \texttt{true} \\
 & \texttt{gm\_latent\_stats\_filename} & \texttt{latent\_whitener.pt} \\
 & \texttt{gm\_disable\_random\_sample} & \texttt{true} \\
\midrule
Dynamics model & \texttt{gm\_dynamics\_model} & \texttt{completep\_resnet} \\
 & \texttt{gm\_dynamics\_d\_model} & $384$ \\
 & \texttt{gm\_dynamics\_n\_layers} & $6$ \\
 & \texttt{gm\_dynamics\_n\_heads} & $8$ \\
 & \texttt{gm\_dynamics\_hidden\_dim} & $1{,}024$ \\
 & \texttt{gm\_dynamics\_width\_multiplier} & $3.0$ \\
 & \texttt{gm\_dynamics\_depth\_multiplier} & $3.0$ \\
 & \texttt{gm\_dynamics\_time\_embed} & \texttt{true} \\
 & \texttt{gm\_dynamics\_time\_freqs} & $10$ \\
 & \texttt{gm\_dynamics\_t\_encoder\_type} & \texttt{logsnr\_sinusoidal} \\
 & \texttt{gm\_dynamics\_x\_encoder\_type} & \texttt{linear} \\
 & \texttt{gm\_dynamics\_activation} & \texttt{gelu} \\
 & \texttt{gm\_dynamics\_norm\_method} & \texttt{ln} \\
 & \texttt{gm\_dynamics\_dropout} & $0.0$ \\
 & \texttt{init\_std} & $0.02$ \\
\midrule
Flow-matching objective & \texttt{gm\_flow\_path} & \texttt{ot} \\
 & \texttt{gm\_t\_lower\_eps} & $10^{-5}$ \\
 & \texttt{gm\_t\_upper\_eps} & $10^{-5}$ \\
 & \texttt{gm\_t\_norm\_scale\_cutoff} & $1$ \\
 & \texttt{gm\_self\_condition\_prob} & $-1$ \\
 & \texttt{gm\_ae\_total\_weight} & $-1.0$ \\
 & \texttt{gm\_loss\_type} & \texttt{l2} \\
\midrule
Optimization / training & \texttt{precision} & \texttt{bf16-mixed} \\
 & \texttt{compile\_model} & \texttt{true} \\
 & \texttt{lr} & $1\times10^{-4}$ \\
 & \texttt{lr\_end} & $1\times10^{-7}$ \\
 & \texttt{lr\_scheduler} & \texttt{cosine\_with\_lr\_end} \\
 & \texttt{lr\_interval} & \texttt{step} \\
 & \texttt{power} & 1 \\
 & \texttt{warmup\_proportion} & $0.001$ \\
 & \texttt{max\_steps} & $1{,}000{,}000$ \\
 & \texttt{num\_decay\_steps} & $-1$ \\
 & \texttt{batch\_size} & $1{,}024$ \\
 & \texttt{eval\_batch\_size} & $1{,}024$ \\
 & \texttt{accumulate\_grad\_batches} & $1$ \\
 & \texttt{optimizer} & \texttt{adamw} \\
 & \texttt{adam\_betas} & $(0.9, 0.98)$ \\
 & \texttt{adam\_eps} & $10^{-8}$ \\
 & \texttt{weight\_decay} & $0.01$ \\
 & \texttt{gradient\_clip\_val} & $5$ \\
\bottomrule
\end{tabular}
\end{table}


\subsection{EF-TALFM: property-conditioned generation}\label{appendix:hyperparameter_talfm_property_conditioned}
\begin{table}[H]
\centering 
\renewcommand{\arraystretch}{1.12}
\caption{Hyperparameters for PCQM4Mv2 property-supervision enabled EF-TAVAE training.}
\label{tab:deltaai-vae-task328}
\begin{tabular}{@{}p{0.25\textwidth}p{0.4\textwidth}p{0.25\textwidth}@{}}
\toprule
Category & Hyperparameter & Value \\ 
\midrule
Model architecture & \texttt{model} & \texttt{transformer-p} \\
 & \texttt{norm\_first} & \texttt{true} \\
 & \texttt{latent\_dim} & $64$ \\
 & \texttt{encoder\_embed\_dim} & $768$ \\
 & \texttt{encoder\_n\_layers} & $12$ \\
 & \texttt{encoder\_n\_heads} & $12$ \\
 & \texttt{encoder\_dim\_feedforward} & $1{,}536$ \\
 & \texttt{encoder\_width\_multiplier} & $6.0$ \\
 & \texttt{encoder\_depth\_multiplier} & $6.0$ \\
 & \texttt{decoder\_embed\_dim} & $768$ \\
 & \texttt{decoder\_n\_layers} & $12$ \\
 & \texttt{decoder\_n\_heads} & $12$ \\
 & \texttt{decoder\_dim\_feedforward} & $3{,}072$ \\
 & \texttt{decoder\_width\_multiplier} & $6.0$ \\
 & \texttt{decoder\_depth\_multiplier} & $6.0$ \\
 & \texttt{decoder\_explicit\_property\_token} & \texttt{true} \\
 & \texttt{disable\_encoder\_context\_condition} & \texttt{true} \\
\midrule
Optimization / training & \texttt{precision} & \texttt{bf16-mixed} \\
 & \texttt{lr} & $7\times10^{-4}$ \\
 & \texttt{lr\_end} & $7\times10^{-6}$ \\
 & \texttt{lr\_scheduler} & \texttt{cosine\_with\_lr\_end} \\
 & \texttt{warmup\_proportion} & $0.0$ \\
 & \texttt{max\_steps} & $1{,}000{,}000$ \\
 & \texttt{num\_decay\_steps} & $-1$ \\
 & \texttt{batch\_size} & $256$ \\
 & \texttt{accumulate\_grad\_batches} & $1$ \\
 & \texttt{optimizer} & \texttt{adamw} \\
 & \texttt{adam\_betas} & $(0.9, 0.98)$ \\
 & \texttt{adam\_eps} & $10^{-8}$ \\
 & \texttt{weight\_decay} & $5\times10^{-3}$ \\
 & \texttt{gradient\_clip\_val} & $5$ \\
\midrule
Loss / regularization & \texttt{pos\_loss\_type} & \texttt{smoothl1-0.001} \\
 & \texttt{position\_scale} & $1.0$ \\
 & \texttt{pos\_weight} & $10$ \\
 & \texttt{seq\_weight} & $1$ \\
 & \texttt{charge\_weight} & $1$ \\
 & \texttt{spinmp\_weight} & $1$ \\
 & \texttt{context\_weight} & $1$ \\
 & \texttt{kl\_weight} & $1\times10^{-4}$ \\
 & \texttt{kl\_weight\_scheduler} & \texttt{linear} \\
 & \texttt{decoder\_property\_molecule\_only\_prob} & $0.6$ \\
 & \texttt{decoder\_property\_latent\_only\_prob} & $0.4$ \\
 & \texttt{property\_molecule\_only\_loss\_weight} & $0.6$ \\
 & \texttt{property\_latent\_only\_loss\_weight} & $0.4$ \\
 & \texttt{vae\_noise\_scale} & $0.05$ \\
 & \texttt{vae\_disable\_learned\_variance} & \texttt{false} \\
 & \texttt{vae\_corruption\_ratio} & $0.0$ \\
 & \texttt{vae\_free\_bits\_per\_dim} & $0.01$ \\
 & \texttt{vae\_sigma\_clamp\_min} & $-1$ \\
 & \texttt{vae\_sigma\_clamp\_max} & $-1$ \\
 & \texttt{vae\_mu\_var\_floor} & $0.01$ \\
 & \texttt{vae\_mu\_var\_floor\_weight} & $0$ \\
 & \texttt{vae\_mu\_cov\_weight} & $0$ \\
 & \texttt{dropout} & $0.0$ \\
 & \texttt{max\_positions} & $64$ \\
\bottomrule
\end{tabular}

\end{table}
For conditional generation with property-supervision is enabled, molecule-condition route is activated with $60\%$ of time, and latent-only route is activated with the rest $40\%$ of time. 
The coordinate, categorical-sequence, charge, spin-multiplicity, context loss weights are $10$, $1$, $1$, $1$, and $1$; molecule-condition route loss weight is $\tau_{\mathrm{mol}}=0.6$, and molecule-condition route loss weight is $1-\tau_{\mathrm{mol}}=0.4$. These settings correspond to the configuration entries in the table above.
\begin{table}[H]
\centering
\renewcommand{\arraystretch}{1.12}
\caption{Hyperparameter setup for PCQM4Mv2 conditional latent flow matching training in EF-TALFM.}
\label{tab:deltaai-lfm-task116}
\begin{tabular}{@{}p{0.25\textwidth}p{0.4\textwidth}p{0.25\textwidth}@{}}
\toprule
Category & Hyperparameter & Value \\
\midrule
Pretrained VAE & \texttt{pretrained\_vae\_subfolder} & \texttt{vae} \\
 & \texttt{pretrained\_vae\_checkpoint} & \texttt{last.ckpt} \\
\midrule
Latent data setup & \texttt{model} & \texttt{transformer-p} \\
 & \texttt{probabilistic\_model} & \texttt{flow\_matching} \\
 & \texttt{context\_dim} & $1$ \\
 & \texttt{gm\_use\_latent\_stats} & \texttt{true} \\
 & \texttt{gm\_latent\_stats\_filename} & \texttt{latent\_scalar.pt} \\
 & \texttt{gm\_disable\_random\_sample} & \texttt{true} \\
\midrule
Dynamics model & \texttt{gm\_dynamics\_model} & \texttt{completep\_resnet} \\
 & \texttt{gm\_dynamics\_d\_model} & $512$ \\
 & \texttt{gm\_dynamics\_n\_layers} & $12$ \\
 & \texttt{gm\_dynamics\_n\_heads} & $8$ \\
 & \texttt{gm\_dynamics\_hidden\_dim} & $1{,}024$ \\
 & \texttt{gm\_dynamics\_width\_multiplier} & $4.0$ \\
 & \texttt{gm\_dynamics\_depth\_multiplier} & $3.0$ \\
 & \texttt{gm\_dynamics\_time\_embed} & \texttt{true} \\
 & \texttt{gm\_dynamics\_time\_freqs} & $32$ \\
 & \texttt{gm\_dynamics\_t\_encoder\_type} & \texttt{logsnr\_sinusoidal} \\
 & \texttt{gm\_dynamics\_x\_encoder\_type} & \texttt{linear} \\
 & \texttt{gm\_dynamics\_y\_encoder\_type} & \texttt{mlp} \\
 & \texttt{gm\_dynamics\_activation} & \texttt{gelu} \\
 & \texttt{gm\_dynamics\_norm\_method} & \texttt{ln} \\
 & \texttt{gm\_dynamics\_dropout} & 0.0 \\
 & \texttt{init\_std} & 0.02 \\
\midrule
Flow-matching objective & \texttt{gm\_flow\_path} & \texttt{ot} \\
 & \texttt{gm\_t\_lower\_eps} & $10^{-3}$ \\
 & \texttt{gm\_t\_upper\_eps} & $10^{-3}$ \\
 & \texttt{gm\_t\_norm\_scale\_cutoff} & 1 \\
 & \texttt{gm\_self\_condition\_prob} & -1 \\
 & \texttt{gm\_ae\_total\_weight} & -1.0 \\
 & \texttt{gm\_loss\_type} & \texttt{l2} \\
\midrule
Optimization / training & \texttt{precision} & \texttt{bf16-mixed} \\
 & \texttt{compile\_model} & \texttt{true} \\
 & \texttt{lr} & $1.4\times10^{-3}$ \\
 & \texttt{lr\_end} & $2.8\times10^{-6}$ \\
 & \texttt{lr\_scheduler} & \texttt{cosine\_with\_lr\_end} \\
 & \texttt{lr\_interval} & \texttt{step} \\
 & \texttt{power} & 1 \\
 & \texttt{warmup\_proportion} & $0.01$ \\
 & \texttt{max\_steps} & $250{,}000$ \\
 & \texttt{num\_decay\_steps} & $-1$ \\
 & \texttt{batch\_size} & $4{,}096$ \\
 & \texttt{eval\_batch\_size} & $4{,}096$ \\
 & \texttt{accumulate\_grad\_batches} & $1$ \\
 & \texttt{optimizer} & \texttt{adamw} \\
 & \texttt{adam\_betas} & $(0.9, 0.98)$ \\
 & \texttt{adam\_eps} & $10^{-8}$ \\
 & \texttt{weight\_decay} & $0.01$ \\
 & \texttt{gradient\_clip\_val} & $5$ \\
\bottomrule
\end{tabular}
\end{table}

\endgroup

\end{document}